\documentclass[preprint,showpacs,amsmath,amssymb,aps]{revtex4}
\usepackage{graphicx}
\begin{document}

\title{New type of chiral motion in even-even nuclei: the $^{138}$Nd case}

\author{A. A. Raduta$^{a),b)}$,  Al. H. Raduta$^{a)}$ and C. M. Petrache$^{c)}$ }

\address{$^{a)}$ Department of Theoretical Physics, Institute of Physics and
  Nuclear Engineering, Bucharest, POBox MG6, Romania}

\address{$^{b)}$Academy of Romanian Scientists, 54 Splaiul Independentei, Bucharest 050094, Romania}

\address{$^{c)}$Centre de Spectrom\'{e}trie de Masse, Universit\'{e} Paris-Sud and CNRS/IN2P3, B\t{a}timent 104-108,F-91405 Orsay, France}

\begin{abstract}The pheomenological Generalized Coherent State Model Hamiltonian is amended with a many body term describing a set of nucleons moving in a shell model mean-field and interacting among themselves with paring, as well as with  a particle-core interaction involving a quadrupole-quadrupole and a hexadecapole-hexdecapole force and a spin-spin interaction. The model Hamiltonian is treated in a restricted space consisting of the core projected states associated to the bands ground, $\beta, \gamma,\widetilde{\gamma}, 1^+$ and $\widetilde{1^+}$ and  two proton aligned quasiparticles coupled to the states of the ground and dipole bands. The chirally transformed particle-core states are also included. The Hamiltonian contains two terms which are not invariant to the chiral transformations relating  the right-handed frame $({\bf J_F}, {\bf J_p}, {\bf J_n})$ and the left-handed ones $(-{\bf J_F}, {\bf J_p}, {\bf J_n})$, $({\bf J_F}, -{\bf J_p}, {\bf J_n})$, 
 $({\bf J_F}, {\bf J_p}, -{\bf J_n})$ where ${\bf J_F}, {\bf J_p}, {\bf J_n}$ are the angular momenta carried by fermions, proton and neutron bosons, respectively. The energies defined with the particle-core states form four chiral bands, two of them being degenerate in the present formalism. Electromagnetic properties of the chiral bands are investigated. Results are compared with the experimental data on $^{138}$Nd.
\end{abstract}

\pacs{21.10.Ky, 21.10.Re, 21.60.Ev}
\maketitle

\section{Introduction}
Many of the nuclear properties  are explored through the  interaction with an electromagnetic field. The electric and magnetic components of the field are used to unveil some  properties of electric and magnetic nature, respectively. As good examples on this line are the scissors like states \cite{LoIu1,LoIu2,LoIu3} and the spin-flip excitations \cite{Zawischa} which were widely treated by various groups. The scissors mode is associated to the angular oscillation of the proton against the neutron system with a total strength  proportional to the nuclear deformation squared, which confirms the collective character of the excitation \cite{LoIu3,Zawischa}. 

Due to this feature it was believed that the magnetic collective properties, in general, show up in deformed systems. However, this is not supported experimentally, since there are the magnetic dipole bands have been observed often in  spherical nuclei. Indeed, there  exists experimental evidence for the magnetic bands in nearly spherical nuclei where the ratio between the moment of inertia and the B(E2) value for exciting the first $2^+$ from the $0^+$ ground state,
${\cal I}^{(2)}/B(E2)$,
takes large values, of the order of 100(eb)$^{-2}MeV^{-1}$ \cite{Frau}. These large values can be consistently explained
by large transverse magnetic dipole moments which induce dipole magnetic transitions, but negligible charge quadrupole moments \cite{Frau}. Indeed, there are several experimental data sets showing that the dipole bands have large values for $B(M1)\sim 1-3 \mu^2_N$, and very small values of $B(E2)\sim 0.1(eb)^2$ (see for example Ref. \cite{Jenkins}). The states are different from the scissors like ones, exhibiting instead a shears character. A system with a large transverse magnetic dipole moment may also consist of a triaxial core to which a proton particle and a  neutron hole are coupled. The maximal transverse dipole momentum is achieved when, for example, $\bf{j}_p$ is oriented along the short axis of the core and $\bf{j}_n$ along the long axis
of the core which rotates around the intermediate axis.   Suppose that the three orthogonal angular momenta form a right-handed frame. If the Hamiltonian describing the interacting system of protons, neutrons and the triaxial core is invariant to the transformation which changes the orientation of one of the three angular momenta, i.e. the right-handed reference frame is transformed to one of a left-handed one, the system exhibits a chiral symmetry. As always happens, such a symmetry is identified when it is broken and consequently for the two reference frames the system acquires distinct energies, otherwise close to each other. Thus, a signature for a chiral symmetry characterizing a triaxial system is the existence of two $\Delta I=1$ bands which are close in energy.  By increasing the total angular momentum, the gradual alignment of $\bf{j}_p$ and $\bf{j}_n$ to the total $\bf{J}$ takes place and a magnetic band is developed \cite{Frau}.

In Refs. \cite{AAR2014,AAR2015} we attempted to investigate another chiral system consisting of one phenomenological core with two components, one for protons and one for neutrons, and two quasiparticles whose total angular momentum ${\bf J_F}$ is oriented along the symmetry axis of the core, due to the particle-core interaction. In the quoted references we proved that states of total angular momentum ${\bf I}$, where the three components mentioned above carry the angular momenta ${\bf J_p}, {\bf J_n}, {\bf J_F}$  which are mutually orthogonal, do exist. Such a configuration seems to be optimal for defining a large transverse magnetic moment that induces large M1 transitions. In choosing the candidate nuclei with chiral features, we were guided by the suggestion \cite{Frau} that triaxial nuclei may favor orthogonality of the aforementioned three angular momenta and therefore may exhibit a large transverse magnetic moment. In the previous publications, the formalism was applied to  $^{192}$Pt, $^{188}$Os and $^{190}$Os, which satisfy the triaxiality signature condition \cite{AAR2014,AAR2015}.

Here we amend the Hamiltonian used in the previous publications by two particle-core terms with the specific feature that the core factors are anharmonic in the quadrupole bosons and have the tensor properties of a quadrupole and hexadecapole operators, respectively. Correspondingly, the other factors are quasiparticle number conserving terms of quadrupole and hexadecapole type.
 The model Hamiltonian is treated in a restricted  space consisting of the projected states of the core describing six collective bands and of a subspace of states spanned by two quasiparticles with the total angular momentum $J$ aligned along the symmetry axis, which are coupled with the states of the ground band to a total angular momentum larger or equal to $J$.
The chiral properties are studied with the $2qp\otimes core$ ground band states. Alternatively, like in Refs. \cite{AAR2014,AAR2015}, we chose another $2qp\otimes core$ subspace by changing the ground band states with those of the magnetic dipole band $1^+$. The formalism is applied to $^{138}$Nd for which some relevant data are available \cite{Petra1,HJLi}.

The work sketched above is developed according to the following plan. In Section 2 we briefly present the main ingredients of the Generalized Coherent State Model (GCSM) which was used to describe the states of the core. The Hamiltonian describing the particle-core interaction is presented in Section 3. The composition of the core and $2qp\otimes core$ states is analyzed in Sections 4 and 5. One identifies states of the $2qp\otimes core$ subspace where the proton $({\bf J_p})$, neutron    $({\bf J_n})$ and fermion    $({\bf J_F})$ angular momenta are mutually orthogonal, which  is a prerequisite of a large transverse dipole momentum. Numerical results for the specific case of $^{138}$Nd are discussed in Section 6, while section 7 is devoted to the final conclusions.

\section{Brief review of the GCSM}
\renewcommand{\theequation}{2.\arabic{equation}}
\setcounter{equation}{0}

The core is described by the GCSM \cite{Rad2}, which is an extension of the Coherent State Model (CSM) \cite{Rad1} for a composite system of protons and neutrons. 
The main ideas underlying the CSM are as follows.
The usual procedure used to describe the excitation energies with a given boson Hamiltonian is to diagonalize it and fix the structure coefficients such that some particular energy levels be reproduced. For a given angular momentum, the lowest levels belong to the ground, gamma and beta bands, respectively. For example, the lowest  state of angular momentum 2, i.e. $2^+_1$, is a ground band state, the next lowest, $2^+_2$, is a gamma band state, while $2^+_3$ belongs to the $\beta $ band. The dominant components of the corresponding eigenstates are one, two and three phonon states. The harmonic limit of the model Hamiltonian yields a multi-phonon spectrum, while by switching on a deforming anharmonicity, the spectrum is a reunion of rotational bands. The correspondence of the two kinds of spectra, characterizing the vibrational and rotational regimes respectively, is realized according to the Sheline-Sakai scheme \cite{Sheline}. In the near vibrational limit a certain staggering is observed for the $\gamma$ band, while in the rotational extreme, the staggering is different. The bands are characterized by the quantum number $K$ which for the axially symmetric nuclei is 0 for the ground and $\beta$ bands, and equal to 2 for $\gamma$ band. The specific property of a band structure consists of that the E2 transition probabilities within a band is much larger that the ones connecting two different bands. For $\gamma$ stable nuclei, the energies of the states heading the $\gamma$ and $\beta$ bands are ordered as $E_{2^{+}_{\gamma}}> E_{0^{+}_{\beta}}$ while for $\gamma$ unstable nuclei the ordering is reversed. A third class of nuclei  exists where $E_{J^{+}_{\gamma}}\approx E_{J^{+}_{\beta}}$, $J$-even.
{\it These are the fundamental features which should be described by the wave functions of any realistic approach}. CSM builds a restricted basis requiring that the states are orthogonal before and after angular momentum projection and, moreover, accounts for the properties enumerated above. If such a construction is possible, then one attempts to define an effective Hamiltonian which is quasi-diagonal in the selected basis. The CSM is, as a matter of fact, a possible solution in terms of quadrupole bosons \cite{Rad1}.

In contrast to the CSM,  within the GCSM \cite{Rad2} the protons are described by quadrupole
proton-like bosons, $b^{\dagger}_{p\mu}$, while the neutrons by quadrupole neutron-like bosons, $b^{\dagger}_{n\mu}$ .
Since one deals with two quadrupole bosons instead of one, one expects 
to have a more flexible model and to find a simpler solution satisfying the restrictions
required by CSM.  The restricted  collective space is defined  by the states describing the three
major bands-ground, beta and gamma- as well as the band  based on
the  isovector state $1^+$. Orthogonality conditions, required for both intrinsic and projected states, are satisfied by the
following 6 functions which generate, by angular momentum projection,
6 rotational bands:
\begin{eqnarray}
|g;JM\rangle&=&N^{(g)}_JP^J_{M0}\psi_g,~~
|\beta;JM\rangle = N^{(\beta)}_JP^J_{M0}\Omega_{\beta}\psi_g,~~
|\gamma ;JM\rangle = N^{(\gamma)}_JP^J_{M2}(\Omega^{\dag}_{\gamma,p,2}+\Omega^{\dag}_{\gamma,n,2})\psi_g,~~
\nonumber\\
|\tilde{\gamma};JM\rangle &=& N^{(\tilde{\gamma})}_JP^J_{M2}(b^{\dag}_{n2}-b^{\dag}_{p2})\psi_g,
|1;JM\rangle = N^{(1)}_JP^J_{M1}(b^{\dag}_nb^{\dag}_p)_{11}\psi_g,\nonumber\\
|{\tilde 1};JM\rangle &=& N^{(\tilde{1})}_JP^J_{M1}(b^{\dag}_{n1}-b^{\dag}_{p1})\Omega^{\dag}_{\beta}\psi_g,
\psi_g = exp[(d_pb^{\dag}_{p0}+d_nb^{\dag}_{n0})-(d_pb_{p0}+d_nb_{n0})]|0\rangle .
\label{figcsm}
\end{eqnarray}
Here, the following notations have been used:
\begin{eqnarray}
\Omega^{\dag}_{\gamma,k,2}&=&(b^{\dag}_kb^{\dag}_k)_{22}+d_k\sqrt{\frac{2}{7}}
b^{\dag}_{k2},~~\Omega^{\dag}_k=(b^{\dag}_kb^{\dag}_k)_0-\sqrt{\frac{1}{5}}d^2_k,~~k=p,n,
\nonumber\\
\Omega^{\dag}_{\beta}&=&\Omega^{\dag}_p+\Omega^{\dag}_n-2\Omega^{\dag}_{pn},~~
\Omega^{\dag}_{pn}=(b^{\dag}_pb^{\dag}_n)_0-\sqrt{\frac{1}{5}}d^2_p,
\nonumber\\
\hat{N}_{pn}&=&\sum_{m}b^{\dag}_{pm}b_{nm},~\hat{N}_{np}=(\hat{N}_{pn})^{\dag},~~
\hat{N}_k=\sum_{m}b^{\dag}_{km}b_{km},~k=p,n.
\label{omegagen}
\end{eqnarray}
$N_{J}^{(k)}$ with $k=g,\beta,\gamma,\tilde{\gamma},1,\tilde{1}$, involved in the wave functions, are normalization factors calculated in terms of some overlap integrals.

Note that for the gamma band there are two candidates which satisfy the requirements of the CSM. One function is symmetric, while the other one is asymmetric with respect to the proton-neutron permutation. In Refs. [9,12] we used alternatively the two versions for the gamma band and we found out that for some nuclei the fitting procedure yielded a better description of the data both for energies and B(E2) values, when the the choice of the asymmetric function was made. In Ref. [9] it was proved that the asymmetric gamma states can be excited from the ground state by the asymmetric component of the electric quadrupole transition operator. The possibility of having two distinct phases for the collective motion in the gamma band has been also considered in Ref. [18] within a different formalism. We don't claim however that this is a general feature for the gamma band, but that for some nuclei the properties specific for asymmetric state may prevail over those characterizing the symmetric gamma states. On the other hand some experiments like $(\alpha,\alpha^{\prime})$ scattering  support the isoscalar structure of the gamma-band states although even this result is not proved to be in general true. For that reason, in this paper we choose the option of symmetric gamma states.

Following the CSM criteria, for the dipole bands we have also two candidate functions, $|1,JM\rangle$ and $|\widetilde{1};JM\rangle$. In Ref. [9] the asymptotic behavior of both states has been considered in the intrinsic reference frame. The important results which came out are: a) the first state is a generalization of the wave functions used by the two rotor [1] and two liquid drops [17]  models; b) the first state corresponds to a lower excitation energy than the second state.
These two properties induced us us to use $|1,JM\rangle$ as model states for the dipole bands studied in the present work.

All calculations performed so far considered equal deformations for protons and neutrons. The deformation parameter for the composite system is:
\begin{equation}
\rho=\sqrt{2}d_p=\sqrt{2}d_n \equiv \sqrt{2}d.
\end{equation}

The projected states defined by Eq. (2.1) describe the essential nuclear properties in the limiting cases the spherical and well deformed systems:
a) They describe degenerate multiphonon states for the vibrational regime, while for large deformation they are identical with the liquid drop prediction in the strong coupling regime.
The connection between the two mentioned extreme regimes is achieved by smoothly varying the parameter $d$ which simulates the nuclear deformation; b) They are mutually orthogonal before and after angular momentum projection and because of that they might be an useful basis set for studying the quadrupole boson Hamiltonian; c) Written in the intrinsic reference frame the projected states are combinations of different $K$ components. The dominant components of the function superpositions have $K=0$ for the ground and beta bands, $K=2$ for the gamma bands and K=1 for the dipole bands. Actually this feature represents a strong support for the band assignments to the model projected states; d) Usually the collective models start with a boson Hamiltonian including anharmonic terms, which is to be diagonalized in a suitable basis for a set of phenomenological parameters which are fixed by fitting some experimental data. Despite the difficulties which might be encountered due to the tedious diagonalization process, the structure of the resulting wave function is simple. The first state $2^+$ is described by a function which has the one-phonon component with a dominant weight, the second $2^+$ by a wave function with the dominant component the two-boson state, etc. Note that our basis exhibits this property and due to this fact it may be considered as obtained by diagonalizing a complex Hamiltonian. By contrast to the above mentioned procedure, the wave functions considered here are infinite series of bosons which results in accounting for high anharmonicity.

According to the above mentioned salient features of the projected states basis, it is desirable to find a boson Hamiltonian which is effective in that basis, i.e. to be quasi-diagonal.
Besides of this restriction, we require to be of the fourth order in bosons and constructed with the rotation invariants of lowest order in bosons.  Since the basis contains both symmetric and asymmetric states with respect to the proton-neutron (p-n) permutation and these are to be approximate eigenstates of the effective Hamiltonian, this should be symmetric against the p-n permutation.

Therefore, we seek  an effective Hamiltonian for which the projected states (\ref{figcsm}) are, at least in a good approximation, eigenstates in the restricted collective space.
The simplest Hamiltonian fulfilling this condition is:
\begin{eqnarray}
H_{GCSM}&=&A_1(\hat{N}_p+\hat{N}_n)+A_2(\hat{N}_{pn}+\hat{N}_{np})+
\frac{\sqrt{5}}{2}(A_1+A_2)(\Omega^{\dag}_{pn}+\Omega_{np})
\nonumber\\
&&+A_3(\Omega^{\dag}_p\Omega_n+\Omega^{\dag}_n\Omega_p-2\Omega^{\dag}_{pn}
\Omega_{np})+A_4\hat{J}^2,
\label{HGCSM}
\end{eqnarray}
with ${\hat J}$ denoting the proton and neutron total angular momentum.
The Hamiltonian given by Eq. (\ref{HGCSM}) has  only one off-diagonal matrix element in the basis (\ref{figcsm}). That is $\langle \beta;JM|H|\tilde{\gamma};JM\rangle$.
However, our calculations show that this affects the energies of $\beta$ and $\tilde{\gamma}$ bands by an amount of a few keV. Therefore, the excitation energies of the six bands are in a  good approximation, given by the diagonal element:
\begin{equation}
E^{(k)}_J=\langle\phi^{(k)}_{JM}|H_{GCSM}|\phi^{(k)}_{JM}\rangle-
\langle\phi^{(g)}_{00}|H_{GCSM}|\phi^{(g)}_{00}\rangle,\;\;k=g,\beta,\gamma,1,\tilde{\gamma},\tilde{1}.
\label{EkJ}
\end{equation}
The analytical behavior of energies and wave functions in the extreme limits of vibrational and rotational regimes have been studied in Refs.
\cite{Rad2,Rad3,Rad4,Lima,4,Iud}.

Since the Hamiltonian is a linear superposition of the $F$ spin algebra generators:
\begin{eqnarray}
F_0&=&\frac{1}{2}(\hat{N}_p-\hat{N}_n),\;F_{+}=\hat{N}_{pn},\; F_{-}=\hat{N}_{np},\;\;\rm{with} \nonumber\\
F_{\pm}&=&F_1\pm F_2,\;\;F_0=F_3.
\end{eqnarray}
it is manifestly that $H$ is not $F$-spin invariant, e.g., $[F_i,H]\ne 0$, for $i=1,2,3.$
Consequently, the eigenstates of $H$ are $F_0$ mixed states. However, one can easily check that the expected value of $F_0$ corresponding to the projected model states is equal to zero.

It is worth to remark that the proposed phenomenological boson Hamiltonian has a microscopic counterpart obtained through the boson expansion procedure from a many body Hamiltonian for a proton and neutron interacting system. In that case the structure coefficients $A_1, A_2, A_3,A_4$ would be analytically expressed in terms of the one- and a two-body interactions matrix elements.

A detailed review of the results obtained with the CSM and GCSM is presented in the recently published book of one of the present authors \cite{AARBook}.

We note that $H_{GCSM}$ comprises a term which is not invariant to the change the sign of either ${\bf J}_p$ or ${\bf J}_p$. For what follows it is useful to  write $H_{GCSM}$ in the form:

\begin{equation}
H_{GCSM}=H^{\prime}_{GCSM}+2A_4{\bf J}_p\cdot {\bf J}_p.
\label{HasGCSM}
\end{equation}

\section{Extension to a particle-core system}
\label{level3}
\renewcommand{\theequation}{3.\arabic{equation}}
\setcounter{equation}{0}
The particle-core interacting system is described by the following Hamiltonian:
\begin{eqnarray}
H&=&H_{GCSM}+\sum_{\alpha}\epsilon_{a}c^{\dag}_{\alpha}c_{\alpha}-\frac{G}{4}P^{\dag}P
\nonumber\\
&-&\sum_{\tau =p,n}X^{h}_{2}\sum_{m}q^{(\tau)}_{2m}\left(b^{\dag}_{\tau,-m}+(-)^mb_{\tau m}\right)(-)^m -X_{sS}{\bf {J}_F}\cdot{\bf {J}_c}\nonumber\\ 
&-&\sum_{J=2,4;\tau=p,n}X^{(an)}_{J}v^{(\tau)}_{J,-m}(-)^m[(b^{\dagger}_{\tau}b^{\dagger}_{\tau})_{J,m}+(\tilde{b}_{\tau}\tilde{b}_{\tau})_{J,m}]
\label{modelH}
\end{eqnarray}
with the notation for the particle multipole operators:
\begin{eqnarray}
q^{(\tau)}_{2m}&=&\sum_{a,b}Q^{(\tau)}_{a,b}\left(c^{\dag}_{j_a}c_{j_b}\right)_{2m},~~
Q^{(\tau)}_{a,b}=\frac{\hat{j}_{a}}{\hat{2}}\langle j_{a}||r^2Y_2||j_b\rangle \nonumber\\
v^{(\tau)}_{Jm}&=&\sum_{a,b}T^{(\tau)}_{a,b;J}\left(c^{\dag}_{j_a}c_{j_b}\right)_{Jm},~~T^{(\tau)}_{a,b;J}=\frac{\hat{j}_{a}}{\hat{J}}\langle j_{a}||r^JY_J||j_b\rangle ,\;J=2,4 
\end{eqnarray}
The core is described by $H_{GCSM}$, while the particle system by  the next two terms standing for a spherical shell model mean-field and pairing interaction of the alike nucleons, respectively. 
The notation $|\alpha\rangle =|nljm\rangle =|a,m\rangle$ is used for the spherical shell model states.
The last three terms, denoted hereafter as $H_{pc}$, express the interaction between the satellite particles and the core. The interaction consists of a quadrupole-quadrupole $qQ$,  a spin-spin $sS$ and an anharmonic $2^J$ pole , with J=2 and 4, denoted by $v_JV_J$  force ,
 respectively, $V_J$ being the anharmonic boson factors which are tensors of rank J. The angular momenta carried by the core and particles are denoted by $\bf{J}_c (= \bf{J}_{p}+\bf{J}_{n})$ and $\bf{J}_F$, respectively. Note that the anharmonic coupling terms are specific to the present paper. The strengths of these interactions denoted by $X^{h}_{2}, X^{an}_J$ ( with J=2,4) and $X_{sS}$ are free parameters.

The mean field plus the pairing term is quasi-diagonalized by means of the Bogoliubov-Valatin transformation.
The free quasiparticle term is $\sum_{\alpha}E_{a}a^{\dag}_{\alpha}a_{\alpha}$, while the $qQ$ and $vV$ interaction preserve  the above mentioned form, with the factors $q_{2m}$ and $v_{Jm}$ changed to:
\begin{eqnarray}
q^{(\tau)}_{2m}&=& \eta^{(-)}_{ab\tau}\left(a^{\dag}_{j_a}a_{j_b}\right)_{2m}+\xi^{(+)}_{ab\tau}\left((a^{\dag}_{j_a}a^{\dag}_{j_b})_{2m}-(a_{j_a}a_{j_b})_{2m}\right),\nonumber\\
v^{(\tau)}_{Jm}&=&\bar{\eta}^{(-)}_{ab\tau;J}\left(a^{\dag}_{j_a}a_{j_b}\right)_{Jm}+\bar{\xi}^{(+)}_{ab\tau;J}\left((a^{\dag}_{j_a}a^{\dag}_{j_b})_{Jm}-(a_{j_a}a_{j_b})_{Jm}\right),\;\; \rm{where}
\nonumber\\
\eta^{(-)}_{ab\tau}&=&\frac{1}{2}Q^{(\tau)}_{ab}\left(U_aU_b-V_aV_b\right),\;\;
\xi^{(+)}_{ab\tau}=\frac{1}{2}Q^{(\tau)}_{ab}\left(U_aV_b+V_aU_b\right), \nonumber\\
\bar{\eta}^{(-)}_{ab\tau;J}&=&\frac{1}{2}T^{(\tau)}_{ab;J}\left(U_aU_b-V_aV_b\right),\;\;
\bar{\xi}^{(+)}_{ab\tau;J}=\frac{1}{2}T^{(\tau)}_{ab;J}\left(U_aV_b+V_aU_b\right).
\end{eqnarray}
The notation $a^{\dagger}_{jm}$ $(a_{jm})$ is used for the quasiparticle creation (annihilation) operator.
We restrict the single-particle space to a proton single-$j$ state. 
In the space of the particle-core states we, therefore, consider the basis defined by:
\begin{eqnarray}
|BCS\rangle\otimes|g;JM\rangle&,&\nonumber\\
\Psi^{(2qp;c)}_{JI;M}&=&N^{(2qp;c)}_{JI}\sum_{J'=even}C^{J\;J'\;I}_{J\;0\;J}\left(N^{(g)}_{J'}\right)^{-1}\left[(a^{\dag}_ja^{\dag}_j)_J|BCS\rangle\otimes|g;J'\rangle \right]_{IM},
\label{basis1}\\
|BCS\rangle\otimes|1;JM\rangle&,&\nonumber\\
\Psi^{(2qp;J1)}_{JI;M}&=&N^{(2qp;J1)}_{JI}\sum_{J'=even}C^{J\;J'\;I}_{J\;1\;J+1}\left(N^{(1)}_{J'}\right)^{-1}\left[(a^{\dag}_ja^{\dag}_j)_J|BCS\rangle\otimes|1;J'\rangle \right]_{IM}.
\label{basis2}
\end{eqnarray}
where $|BCS\rangle$ denotes the quasiparticle vacuum, while $N_{JI}^{(2qp;c)}$ and $N_{JI}^{(2qp;J1)}$ are the projected state norms. 

{\it Why these bases states are favored in describing the chiral properties of some nuclei?} The states describing the core might be any of the eigenstates of $H_{GCSM}$. The lowest bands are energetically preferred and thus a good candidate is the ground-state band. On the other hand, in Ref. \cite{AAR2014} it was shown that the states of the isovector dipole band are also  good candidates since the core contribution to the $M1$ transitions would be substantial. Since the basis (\ref{basis2}) was described in detail in Refs. \cite{AAR2014,AAR2015}, we restrict our presentation to the basis (\ref{basis1}).

According to the definition (2.1) the unprojected ground state is a product of two coherent states associated to the proton and neutron systems, respectively. The angular momentum projected component from each factor state is a 
$K=0$ function, i.e, in the intrinsic coordinates system the $K=0$ component of the wave function prevails over the other ones. Therefore, the angular momenta  carried by protons and neutrons  are lying in a plane perpendicular on the intrinsic symmetry axis. In the state $0^+$ belonging to the ground band of the core, the two angular momenta are anti-aligned. When the angular momentum of the core increases the proton and neutron angular momenta tend to align to each other, their relative angle gradually decreases and finally vanishes for high total angular momentum. We recognize a shears-like motion of the proton and neutron angular momenta of the core. On the path to the mentioned limit the angle reaches the value of $\pi/2$,
 which is necessary to have an optimal configuration for the magnetic dipole transition.
On the other hand if the two quasiparticle state has a maximum projection of the angular momentum on the $z$ axis, which is chosen to coincide with the symmetry axis, the total quasiparticle angular momentum is perpendicular to each of the core angular momentum, realsizing, thus, the dynamical chiral symmetry. In Ref. \cite{AAR2015}, the calculations were performed for $(a^{\dagger}_ja^{\dagger}_j)_{JK}$ with $K=0,2,...2j-1$. The result was that $K=2j-1$  yields the maximal magnetic dipole transition probabilities. The orientation of the quasiparticle angular momenta is specific to the hole-like protons since the corresponding wave function overlap with the density distribution is maximal. Actually, the $qQ$ interaction  favors the motion of particles on orbits close to the equator of the density ellipsoid.
It is noteworthy to make a degression concerning the particle-core subspace. We have proved in several publications that projecting the angular momentum from a spherical shell-model state of maximum $m$-projection times the unprojected ground-state, one can define a  basis for the particle-core space which, moreover, can be also used as a single particle basis \cite{RB2014}. The same arguments work for the space of two particles-core states in the laboratory frame. This basis has the advantage of having $K=2j-1$ as the dominant component which is suitable for the description of large M1 transitions. 
 
In conclusion, this basis is optimal in order to describe a composite system which rotates around an axis not situated  in any principal planes of the density distribution ellipsoid.

It is known that in the nuclei of the A=130-140 mass region, which are $\gamma$-soft, the valence protons occupy the lower half of the $h_{11/2}$ orbital driving the nucleus to a prolate shape (
$\gamma\sim 0^{\circ}$), while the valence neutrons occupy the upper half of the $h_{11/2}$ orbital, which favors an oblate shape ($\gamma\sim 60^{\circ}$). One expects, therefore, coexisting prolate and oblate minima for the potential energy. For nuclei with $Z\sim 60$ the $h_{11/2}$[505]9/2 and $h_{11/2}$[505]11/2 orbitals are strongly down sloping in energy on the oblate side ($\varepsilon_2<0$ of the Nilsson diagram and may aslo contribute to the stability of the oblate shape. Indeed, collective oblate bands built on or involving these single high-$\Omega$$h_{11/2}$ orbitals have been observed to low spin ($11/2^-$) in light iodine nuclei \cite{Lia90} and at medium spins in $^{136}$Ce \cite{Paul90} and references therein.

We note that the Hamiltonian $H$ is not invariant to the chiral transformations, defined by changing the sign of one of the angular momenta ${\bf J_F}, {\bf J_p}$ and ${\bf J_n}$. It is worthwhile to  write $H$ as a sum of a chiral invariant term $H^{\prime}$ and the non-invariant terms:
\begin{eqnarray}
H&=& H^{\prime} +2A_4{\bf J_p}\cdot {\bf J_n}-X_{sS}{\bf J_{F}}\cdot {\bf J_{c}}=H^{\prime\prime}-X_{sS}{\bf J_{F}}\cdot {\bf J_{c}}\nonumber\\
&=&H_1+2A_4{\bf J_p}\cdot {\bf J_n}
\label{HHprimH1}
\end{eqnarray} 
The chiral properties will be studied alternatively with the basis $(\ref{basis1})$ and $(\ref{basis2})$

\section{Study of H in the basis $(\ref{basis1})$}

\subsection{ Angular momenta composition of the projected states}
For a state characterizing the ground band, the results concerning the average values of the angular momenta carried by protons and neutrons respectively, are:
\begin{equation}
\widetilde{J}^{(g)}_{\tau; J}( \widetilde{J}^{(g)}_{\tau; J}+1)\equiv\langle g;JM|\hat{\bf J}^{2}_{\tau}|g;JM\rangle =\left(N^{(g)}_{J}\right)^2\sum_{J_pJ_n} \left(N^{(g)}_{J_p}\right)^{-2}\left(N^{(g)}_{J_n}\right)^{-2}
J_{\tau}(J_{\tau}+1)\left(C^{J_pJ_nJ}_{0\;\;0\;\;0}\right)^2,\;\tau=p,n.
\end{equation}
Note that $\widetilde{J}^{(g)}_{pJ}=\widetilde{J}^{(g)}_{nJ}$, since the proton and neutron deformations are equal to each other. Using the above equation, the angle between the two angular momenta associated to the proton and neutron subsystems is readily obtained:
\begin{equation}
\cos({\bf J}_p,{\bf J}_n)=\frac{J(J+1)-\widetilde{J}^{(g)}_{pJ}(\widetilde{J}^{(g)}_{pJ}+1)-\widetilde{J}^{(g)}_{nJ}(\widetilde{J}^{(g)}_{nJ}+1)}
{2\sqrt{\widetilde{J}^{(g)}_{pJ}(\widetilde{J}^{(g)}_{pJ}+1)\widetilde{J}^{(g)}_{nJ}(\widetilde{J}^{(g)}_{nJ}+1)}}.
\end{equation}
Similarly the angular momenta composition of the $2qp\otimes core$ states is specified by the following relations:
\begin{eqnarray}
\widetilde{J}^{(2qp,c)}_{\tau;JI}(\widetilde{J}^{(2qp,c)}_{\tau;JI}+1)&\equiv& \langle\Psi^{(2qp,c)}_{JI}|\hat{{\bf J}}^{2}_{\tau}|\Psi^{(2qp,c)}_{JI}\rangle\nonumber\\
&=&\left(N^{(2qp;c)}_{JI}\right)^2\sum_{J'}2\left(C^{JJ'I}_{J0J}\right)^2\left(N^{(g)}_{J'}\right)^{-2}\widetilde{J}^{(g)}_{\tau;J'}(\widetilde{J}^{(g)}_{\tau;J'}+1),\;\tau=p,n,\nonumber\\
\widetilde{J}^{(2qp,c)}_{pn;JI}(\widetilde{J}^{(2qp,c)}_{pn;JI}+1)&\equiv& \langle\Psi^{(2qp,c)}_{JI}|(\hat{{\bf J}}_{p}+\hat{{\bf J}}_{n})^2|\Psi^{(2qp,c)}_{JI}\rangle\nonumber\\
&=&\left(N^{(2qp;c)}_{JI}\right)^2\sum_{J'}2\left(C^{JJ'I}_{J0J}\right)^2\left(N^{(g)}_{J'}\right)^{-2}\widetilde{J}^{(g)}_{pn;J'}(\widetilde{J}^{(g)}_{pn;J'}+1).
\end{eqnarray}
Here we used the notation $\hat{\bf J}_{pn}=\hat{\bf J}_{p}+\hat{\bf J}_{n}$ and $\widetilde{J}^{(g)}_{pn;J'}$ for the average of $\hat{J}_{pn}$ with the ground band state $|g;JM\rangle$.
The angle between the proton and neutron angular momenta can be obtained from the equation:
\begin{equation}
\cos({\bf J}_p,{\bf J}_n)^{(2qp,c)}_{JI}=\frac{\widetilde{J}^{(2qp,c)}_{pn;JI}(\widetilde{J}^{(2qp,c)}_{pn;JI}+1)-\widetilde{J}^{(2qp,c)}_{p;JI}(\widetilde{J}^{(2qp,c)}_{p;JI}+1)-
\widetilde{J}^{(2qp,c)}_{n;JI}(\widetilde{J}^{(2qp,c)}_{n;JI}+1)}
{2\sqrt{\widetilde{J}^{(2qp,c)}_{p;JI}(\widetilde{J}^{(2qp,c)}_{p;JI}+1)\widetilde{J}^{(2qp,c)}_{n;JI}(\widetilde{J}^{(2qp,c)}_{n;JI}+1)}}.
\end{equation}
\begin{figure}[h!]
\begin{center}
\includegraphics[width=0.4\textwidth]{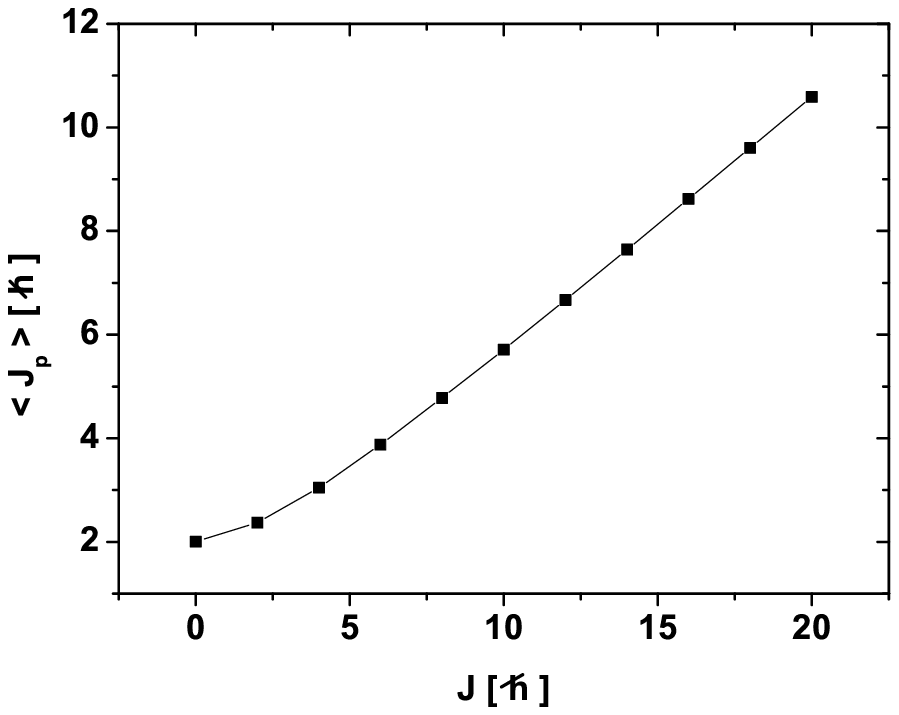}\includegraphics[width=0.4\textwidth]{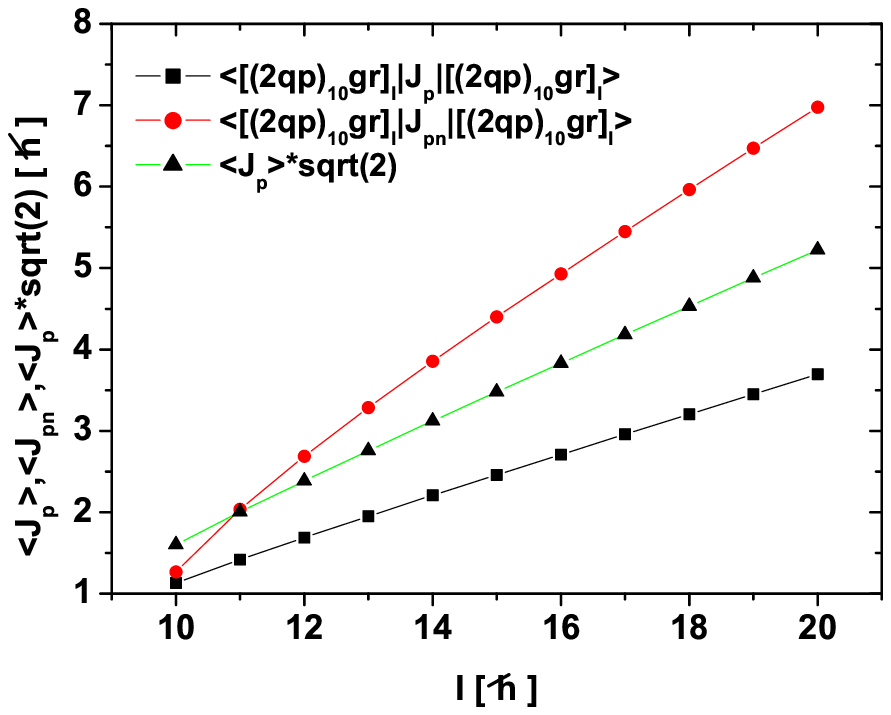}
\end{center}
\begin{minipage}[h]{7cm}
\caption{\scriptsize{ The average value of the proton angular momentum $<J_p>$, corresponding to a ground-state band as function of the angular momentum.}}
\end{minipage}\ \
\begin{minipage}[h]{7.5cm}
\caption{\scriptsize{The average value of the proton angular momentum $<J_p>$ and  the total proton plus neutron  angular momentum, $<J_{pn}>$, for the 2qp $\otimes$ core-ground band. Also $<J_p>\sqrt{2}$ is represented.}}
\end{minipage}
\end{figure}
The average values of the angular momenta in a ground band state as well as in a state of the $2qp\otimes core$ system are plotted as function of the total angular momentum in Figs. 1 and 2.
In Fig. 2 we also show the values of $\widetilde{J}^{(g)}_{p;JI}\sqrt{2}$ to be compared with $\widetilde{J}^{(g)}_{pn;JI}$. When the two curves intersect each other, the angular momenta carried by protons and neutrons respectively, are orthogonal. Actually, the distance between the two curves represents a measure for the deviation from the orthogonality geometry.
The angles between the proton and neutron angular momenta in a ground band state as well as in a  $2qp\otimes core$ state are plotted as function of the state total angular momentum in Figs. 3 and 4 respectively.
The mechanism responsible for the orthogonality of proton and neutron angular momenta is determined by the increasing rotation frequency. Indeed, in the ground state the core's proton and neutron angular momenta are anti-aligned, leading to a total angular momentum  equal to zero. Increasing the total angular momentum and therefore the rotational frequency, the angle between the  two angular momenta becomes smaller than $\pi$ and for some critical values of frequency it reaches the value of $\frac{\pi}{2}$. The angle between the core proton and neutron angular momenta decreases smoothly and slowly tends to zero for very high angular momenta.

The situation when the core proton and neutron, as well as the fermion angular momenta are mutually orthogonal is optimal in determining the dipole transition strength.
When the model Hamiltonian is invariant to any chiral transformation one says that it exhibits a chiral symmetry and therefore any state of the $2qp\otimes core$ and its transformed state are degenerate. Breaking gently the chiral symmetry, this degeneracy is removed and as a result two $\Delta I=1$ quasi-degenerate bands show up. Actually, the chiral symmetry is detected just by identifying the chiral bands partners. In other words, the symmetry is indirectly figured out when this is broken. 
Breaking the chiral symmetry leads to a nuclear phase transition, one phase being characterized by the chiral symmetry, while the other one without such a symmetry. Usually, a phase transition
is evidenced by the behavior of an order parameter. For the phase transition mentioned above a possible  order parameter could be \cite{Grod}:
\begin{equation}
\epsilon = \frac{({\bf J_p}\times {\bf J_n})\cdot{\bf J_F}}{\sqrt{J_p(J_p+1)J_n(J_n+1)J_F(J_F+1)}}.
\end{equation}
The extreme values of $\epsilon$ are 1 when the three vectors are mutually orthogonal and 0 when they are planar.
This quantity is plotted as function of the total angular momentum in Fig 5. Note that for I=11, 12 the angular momenta carried by the three components are orthogonal.

\subsection{Chiral features within the basis (\ref{basis1})}

 Suppose that the mentioned orthogonal angular momenta form a right-handed frame denoted by $F_1$. Acting on $F_1$ with the chiral transformation ${\bf J_F}\to -{\bf J_F}$, one obtains the left-handed frame $F_2$. This transformation, conventionally denoted by $C_{12}$, does not commute with $H$, due to the spin-spin term, $sS$.

\begin{figure}[h!]
\begin{center}
\includegraphics[width=0.4\textwidth]{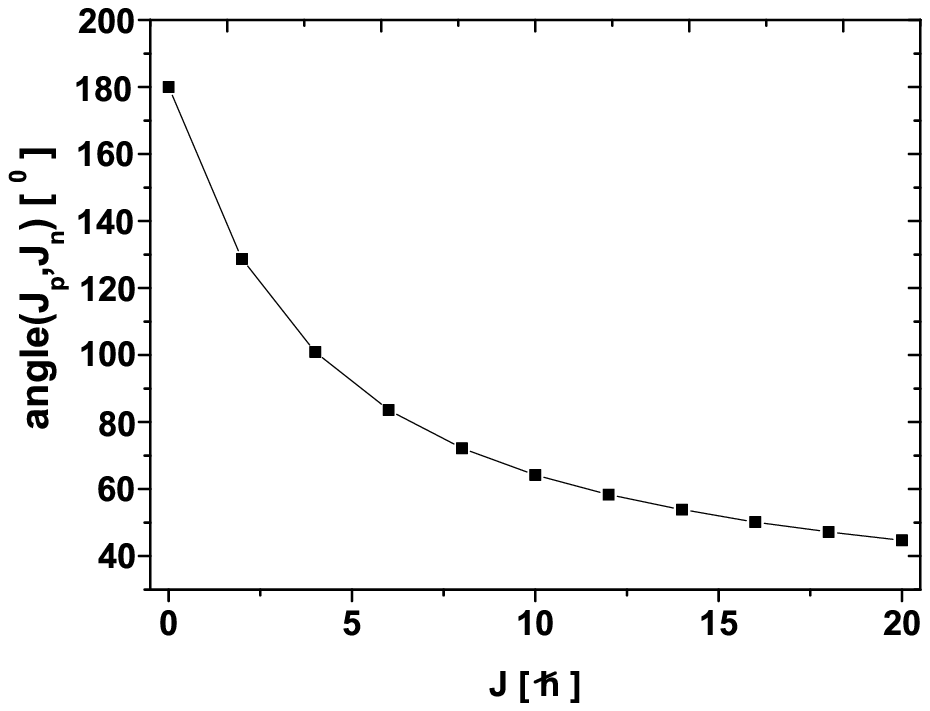}\includegraphics[width=0.4\textwidth]{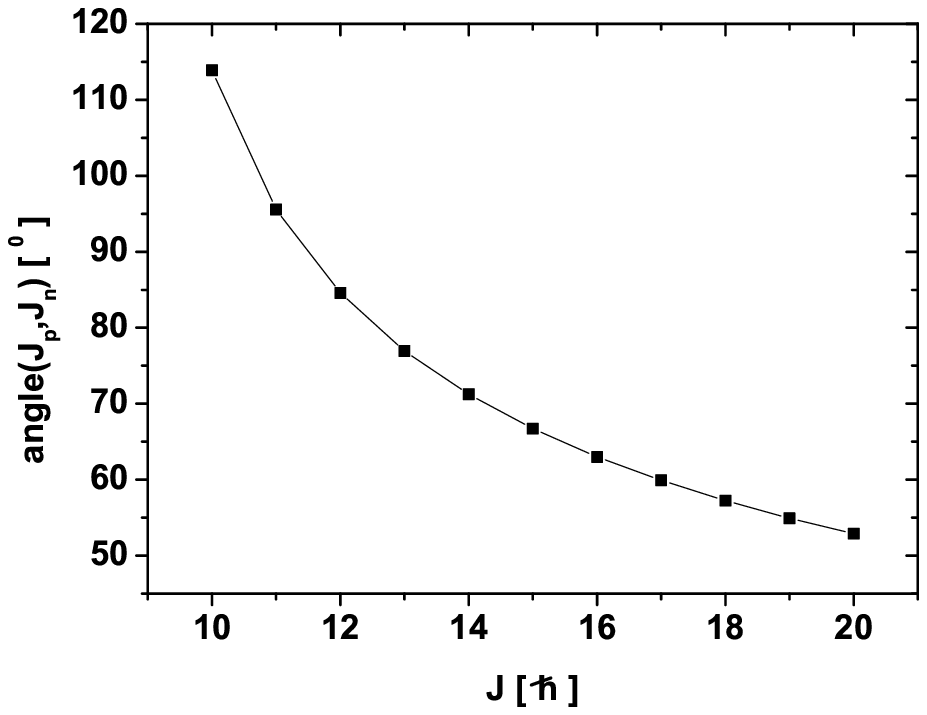}
\end{center}
\begin{minipage}[h]{7cm}
\caption{\scriptsize{ The angle between the angular momenta of the core protons and neutrons in a ground-state band.}}
\end{minipage}\ \
\begin{minipage}[h]{7.5cm}
\caption{\scriptsize{The angle between the angular momenta of the core protons and neutrons in a state of the 2qp$ \otimes$ core-ground state  band. }}
\end{minipage}
\end{figure}

\begin{figure}[h!]
\begin{center}
\includegraphics[width=0.7\textwidth]{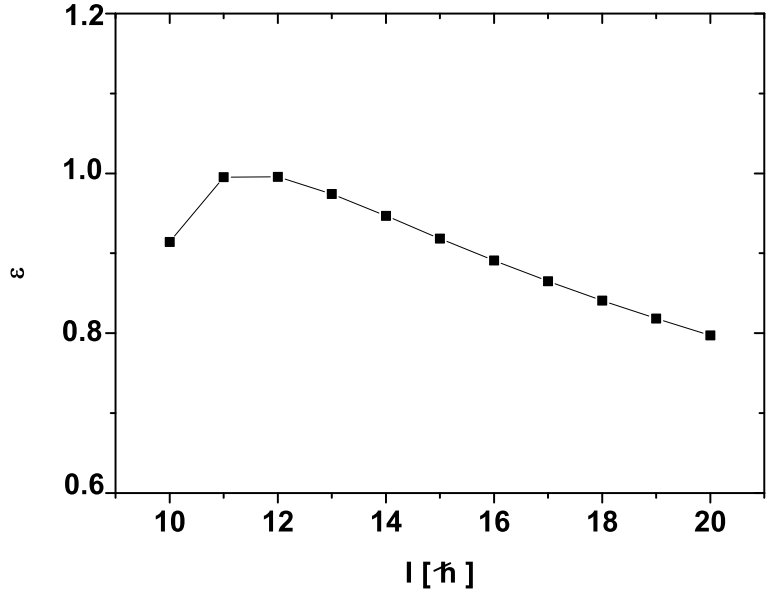}
\end{center}
\caption{\scriptsize{ The order parameter $\epsilon$ as function the state total  angular momentum.}}
\end{figure}
However it  anti-commutes with the spin-spin term:
\begin{equation}
\{-X_{sS}{\bf J_F}\cdot {\bf J_c, C_{12}}\}=0.
\end{equation}
If $|\psi\rangle$ is an eigenstate of $-X_{sS}{\bf J_F}\cdot {\bf J_c}$ corresponding to the eigenvalue $\lambda$ then the transformed function $C_{12}|\psi\rangle$
is also eigenfunction corresponding to the eigenvalue $-\lambda$. One eigenfunction of the spin-spin interaction is $\Psi^{(2qp;c)}_{JI;M}$ with the eigenvalue
\begin{equation}
\lambda_{JI}=-X_{sS}\left(N^{(2qp;c)}_{JI}\right)^2\sum_{J^{\prime}}\left(C^{J\;J^{\prime}\;I}_{J\;0\;J}\right)^2\left(N^{(g)}_{J^{\prime}}\right)^{-2}\left[I(I+1)-J(J+1)-J^{\prime}(J^{\prime}+1)\right].
\end{equation}
Obviously, the spectrum of the spin-spin interaction has the chiral property since a part of it is the mirror image, with respect to zero, of the other one.

Let us see what happens with the whole Hamiltonian (3.6). It is easy to show that the following equations approximatively hold:
\begin{eqnarray}
\left[H^{\prime\prime}-X_{sS}{\bf J_{F}}\cdot {\bf J_{c}}\right]\Psi^{(2qp;c)}_{JI;M}&=&\left[\langle \Psi^{(2qp;c)}_{JI;M}|H^{\prime\prime}|\Psi^{(2qp;c)}_{JI;M}\rangle+\lambda_{JI}\right]|\Psi^{(2qp;c)}_{JI;M}\rangle , \nonumber\\
\left[H^{\prime\prime}-X_{sS}{\bf J_{F}}\cdot {\bf J_{c}}\right]C_{12}|\Psi^{(2qp;c)}_{JI;M}\rangle&=&\left[\langle \Psi^{(2qp;c)}_{JI;M}|H^{\prime\prime}|\Psi^{(2qp;c)}_{JI;M}\rangle-\lambda_{JI}\right]
C_{12}|\Psi^{(2qp;c)}_{JI;M}\rangle .
\end{eqnarray}
Therefore the Hamiltonian $H$ exhibits also the chiral feature, since a part of the spectrum is the mirror image of the other, with respect to an intermediate spectrum given by averaging $H^{\prime\prime}$ with the function $|\Psi^{(2qp;c)}_{JI;M}\rangle$.

Similar considerations can also be applied to the Hamiltonian $H$ and to the basis of $|\Psi^{(2qp;c)}_{JI;M}\rangle $ and $C_{13}|\Psi^{(2qp;c)}_{JI;M}\rangle$, where $C_{13}$ denotes the chiral transformation which changes ${\bf J_p}$ to ${-\bf J_p}$. Since the transformed Hamiltonian with the chiral transformation ${\bf J_n}\to -{\bf J_n}$, denoted hereafter by $C_{14}$, is identical with the one corresponding to the transformation
${\bf J_p}\to -{\bf J_p}$, the two transformed Hamiltonians have identical spectra.

Summarizing the above results, one can say
that the spectrum of $H$ within this restricted space $|\Psi^{(2qp;c)}_{JI;M}\rangle$ forms a  chiral band denoted, for simplicity, with $T1$.  The eigenvalues of $H$  obtained by averaging $H$ with the transformed wave function $C_{12} |\Psi^{(2qp;c)}_{JI;M}\rangle$, forms the chiral partner band denoted by $T2$. Another partner band of $T1$ is $T3$  corresponding to the chiral transformed functions
$C_{13}|\Psi^{(2qp;c)}_{JI;M}\rangle$. Also the partner band of $T1$, denoted hereafter with $T4$, obtained by averaging $H$ with $C_{14}|\Psi^{(2qp;c)}_{JI;M}\rangle $ is identical with $T3$.
Note that the symmetry generated by the transformations $C_{13}$ and $C_{14}$ are broken by two terms, namely the spin-spin particle-core interaction and the rotational term $2A_4{\bf J_p}.{\bf J_n}$ involved in $H_{GCSM}$. The latter term is ineffective in a state where the angular momenta ${\bf J_p},{\bf J_n}$  are orthogonal. Since the wave-function $|\Psi^{(2qp;c)}_{JI;M}\rangle$ is symmetric with respect to the proton-neutron permutation, the average values of the $sS$ term with the transformed functions $C_{13}|\Psi^{(2qp;c)}_{JI;M}\rangle $ and 
$C_{14}|\Psi^{(2qp;c)}_{JI;M}\rangle $ are vanishing. Therefore, the degenerate bands $T3$ and $T4$ are  essentially determined by the symmetry breaking term generated by $A_4\hat{J}^2$, i.e. -4$A_4{\bf J_p}\cdot {\bf J_n}$.   Concluding, there are four chiral partner bands $T1, T2, T3, T4$,  obtained  with $H$ and the wave functions $|\Psi^{(2qp;c)}_{JI;M}\rangle$,  $C_{12}|\Psi^{(2qp;c)}_{JI;M}\rangle$, $C_{13}|\Psi^{(2qp;c)}_{JI;M}\rangle$, $ C_{14}|\Psi^{(2qp;c)}_{JI;M}\rangle$, respectively. The four chiral bands will be quantitatively studied in  section VI.

The chiral transformation can always be written as a product of a rotation with an angle equal to $\pi$ and a time reversal operator. The rotation is performed around one of the axes defined by 
${\bf J_F}$, ${\bf J_p}$, ${\bf J_n}$  which, in the situation when they are an orthogonal set of vectors, coincide with the axes of the body fixed frame. Since  the Hamiltonian has terms which are linear in the rotation generators mentioned above, it is not invariant with respect to these rotations. $H$ is however invariant to the rotations in the laboratory frame, generated by the components of the total angular momentum, $ \bf J_F+\bf J_p+\bf J_n $.  The fingerprints of such an invariance can be found also in the structure of the wave functions describing the eigenstates of $H$ in the laboratory frame. This can be easily understood having in mind the following aspects. The proton and neutron angular momenta of the core are nearly perpendicular vectors in a plane perpendicular to the symmetry axis, in a certain spin interval. However, we cannot state that ${\bf J}_p$ is oriented along the $x$ or $y$ axes. In the first case the intrinsic reference frame would be right-handed, while in the second situations it is left-handed. In other words the wave function must comprise right- and left-handed components which are equally probable. 
Their weights are then either identical or equal in magnitude but of opposite sign. Since the transformation $C_{13}$ changes the direction of ${\bf J_p}$, it will change the left- to  the right-handed component and vice versa. It results that the states of the basis (\ref{basis1}) are eigenstates of $C_{13}$. Similar reasoning is valid also for the transformation $C_{12}$, the component corresponding to the orientation of ${\bf J_F}$ along the $z$ axis being equally probable with the component with ${\bf J_F}$  having an opposite direction.  We remark that by the transformation 
$C_{12}$, the wave function $|\Psi^{(2qp;c)}_{JI;M}\rangle$ goes to:
\begin{equation}
|\Psi^{(2qp;c)}_{-J,I;M}\rangle=(-)^I|\Psi^{(2qp;c)}_{JI;M}\rangle.
\end{equation} 
\section{The study of the basis (\ref{basis2})}
Here we formulate a schematic model which is more easy to handle. The basis (\ref{basis2}) has been used in Refs. \cite{AAR2014,AAR2015} in connection with the triaxial isotopes $^{192}$Pt and 
$^{188,190}$Os. Therein, the angular momentum composition of the involved projected states was described in detail. Moreover, the particle-core Hamiltonian was similar to the one used in the present paper with the difference that there the particle-core terms with anharmonic boson factor were missing. Here, we shall use a very simple Hamiltonian, which in the quasiparticle representation looks like:
\begin{equation}
H_2=H^{\prime}_{GCSM}+2A_4{\bf J_p}\cdot {\bf J_n}+\sum_{\alpha}E_{a}a^{\dagger}_{\alpha}a_{\alpha} -X_{sS}{\bf {J}_F}\cdot{\bf {J}_c}.
\end{equation}
We shall use the same notations as in the previous sections. Thus, by the transformations $C_{12},C_{13},C_{14}$, the right-handed intrinsic reference frame $F_1$ is changed to the left-handed frames $F_2$, $F_3$ and $F_4$ respectively.
Correspondingly, the wave functions  $|\Psi^{(2qp;J1)}_{JI;M}\rangle$  are transformed to $C_{12}|\Psi^{(2qp;J1)}_{JI;M}\rangle$, $C_{13}|\Psi^{(2qp;J1)}_{JI;M}\rangle$ and $ C_{14}|\Psi^{(2qp;J1)}_{JI;M}\rangle$, respectively. Each of these functions generates a rotational band defined by the corresponding average values of $H$. Thus, the band denoted  by $B_1$ and its chiral partners $B_2$, $B_3$ and $B_4$ show up. The proof that the bands have, indeed, a chiral character goes identically with that used for the different Hamiltonian (\ref{modelH}) and the different basis 
(\ref{basis1}).

Since the spin-spin interaction is ineffective in determining the band $B_4$  it is natural to construct its chiral partner also with the  Hamiltonian $H_3$ defined by (3.16), but without the spin-spin interaction, i.e.
\begin{equation}
H_3=H^{\prime}_{GCSM}+2{\bf J_p}\cdot {\bf J_n}+\sum_{\alpha}E_{a}a^{\dagger}_{\alpha}a_{\alpha}.
\end{equation}
Thus, the bands obtained by averaging $H_3$ alternatively with $|\Psi^{(2qp;J1)}_{JI;M}\rangle$ and with $C_{13}|\Psi^{(2qp;J1)}_{JI;M}\rangle$ are denoted by $B_3$ and $B_4$ respectively.
The band obtained with $C_{14}|\Psi^{(2qp;J1)}_{JI;M}\rangle$ is degenerated with the band $B_4$

\section{Numerical results and discussion}

The formalism presented above was applied to the case of $^{138}$Nd. The experimental data were taken from Refs. \cite{Petra1,HJLi}. The reason for which we chose to apply the present formalism to $^{138}$Nd is that it has been proven that this nucleus is triaxial both at low and high spins, through a coherent interpretation of most of the multitude of observed bands. The triaxiality is a prerequisite of chiral motion which is predicted by the present formalism. 
Among the eight observed dipole bands in $^{138}$Nd, there are two bands which could have chiral character: band $D3$ which is suggested as chiral partner of band $D2$ in Ref. [10] within the $TAC$ formalism, and band $D4$ for which two possible configurations of two and four quasiparticles were suggested in Ref. [10], but which has characteristics compatible with the present $GCSM$ formalism which predict chiral  bands of different nature. Indeed, for band $D4$ both the level energies and the transition probabilities are in agreement with the present formalism. Firstly, the band-head energy of band $D4$ is the lowest among the eight observed dipole bands and is in agreement with the energy calculated by $GCSM$ using realistic assumptions for the model parameters. Secondly, the band is composed of only dipole transitions, which is in agreement with the prediction of the present formalism. Thirdly, a possible chiral partner has been identified. All these features which will be discussed in the following in this section induced us to propose band $D4$ as candidate for the new chiral mode. The collective states from the ground, $\beta$ and $\gamma$ bands were described by means of the GCSM. The particle-core term is associated to the protons from $h_{11/2}$ interacting with the core through the harmonic $qQ$ term, a spin-spin term and an anharmonic force of quadrupole-quadrupole plus a hexadecapole-hexadecapole type. The structure coefficients $A_1,A_2,A_3, A_4$ were fixed by fitting the experimental energies of the states $2^+_{g},10^+_{g}, 2^+_{\beta},2^+_{\gamma}$. The deformation parameter $\rho$ was chosen such that an overall agreement is obtained. We interpreted the state of 2.273 MeV
as being the state $2^+_{\beta}$, since it is populated by the Gammow-Teller beta decay of the state $3^+$ from $^{138}$Pm \cite{Desla}. As shown in Figs. 6 and 7,  a reasonable agreement with the experimental data is obtained.

As we already mentioned, the constraints of the GCSM formalism are satisfied by two model states for gamma  and two for dipole bands: one  symmetric and  one asymmetric with respect to the proton-neutron permutation for gamma bands and two asymmetric functions for dipole bands.  In our calculations the symmetric states are considered for the gamma band. 
As for the $\beta$ band, we present the energies up to the high spin region, even if  the experimental data are missing. 

Results for the dipole bands $1^+$ and $\widetilde{1}^+$ are shown in Fig. 8. 
\begin{figure}[h!]
\begin{center}
\includegraphics[width=0.4\textwidth]{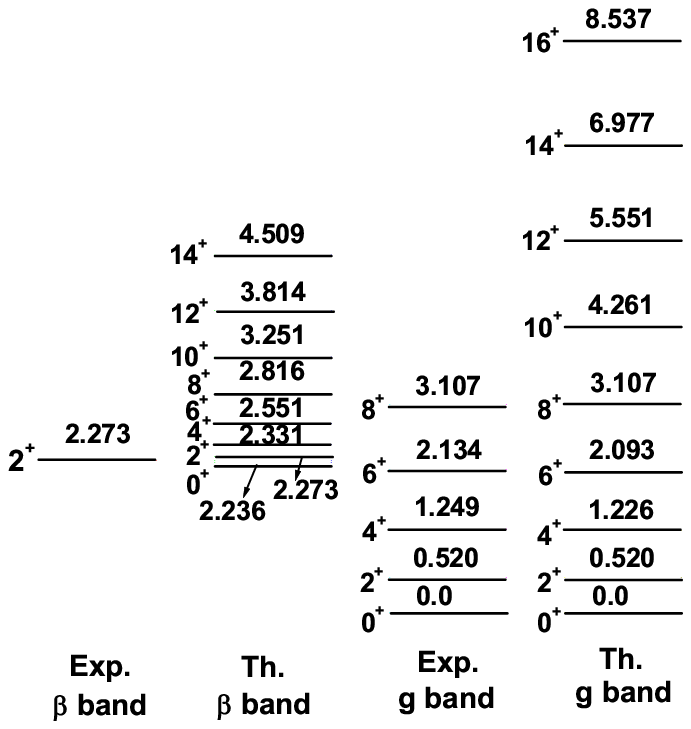}\includegraphics[width=0.4\textwidth]{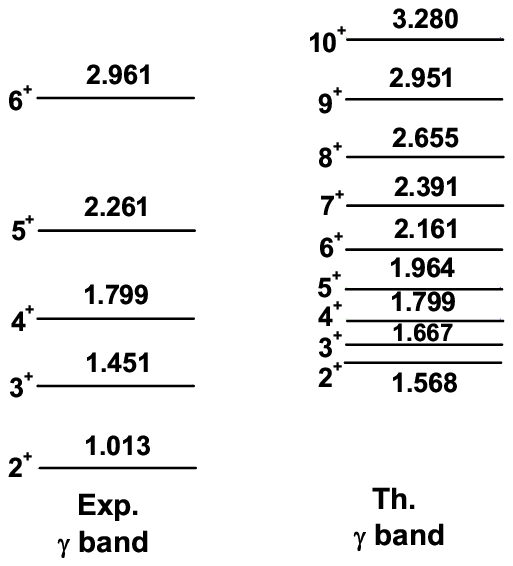}
\end{center}
\begin{minipage}[h]{7cm}
\caption{\scriptsize{ The calculated energies for ground and beta bands are compared with the corresponding experimental values for the ground state band and L1 band.}}
\end{minipage}\ \   
\begin{minipage}[h]{7.5cm}
\caption{\scriptsize{The theoretical values of the $\gamma$ band energies are compared with the corresponding experimental data. }}
\label{Figgama}
\end{minipage}
\end{figure}

\begin{figure}[h!]
\begin{center}
\hspace*{1cm}\includegraphics[width=0.5\textwidth]{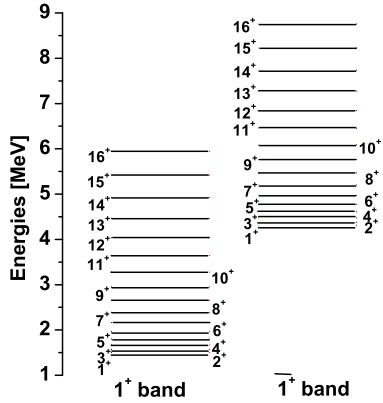}
\end{center}
\caption{\scriptsize{ The calculated excitation energies for the  the dipole band $1^{+}$. Also the energies for the  dipole band $\widetilde{1}^{+}$ are  shown.}}
\end{figure}
The results obtained  with the two options for the 2$qp \otimes$ core states subspace are presented separately, in different subsections.

\subsection{Chiral bands with H, $|\Psi^{(2qp;c)}_{JI;M}\rangle$ and chiral transformed states}

\begin{figure}[h!]
\begin{center}
\includegraphics[width=0.9\textwidth]{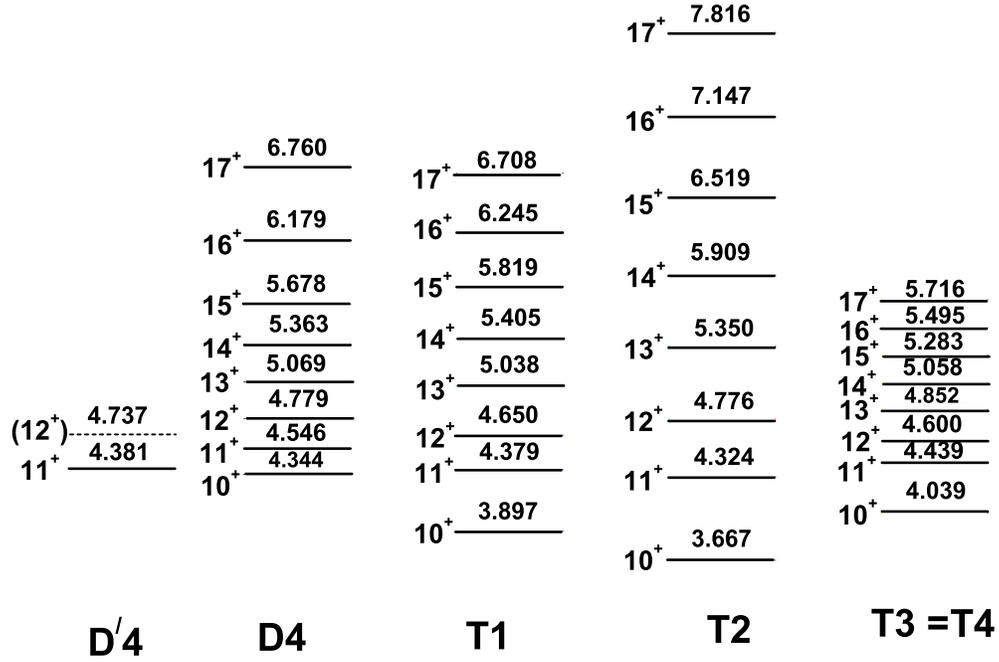}     
\end{center}
\caption{\scriptsize{The chiral band $T1$ is compared with  the experimental band D4. The energies of other three bands, $T2, T3$ and  $T4$,  exhibiting chiral properties are also presented.
The experimental band $D'4$, interpreted as partner band for $D4$, is to be compared with the calculated band $T2$.}}
\label{chirband}
\end{figure}

Finally, the chiral bands defined as the averages of the model Hamiltonian $H$ with the $2qp\otimes core$ states, $C_{13}|\Psi^{(2qp;c)}_{JI;M}\rangle$  and $C_{14}|\Psi^{(2qp;c)}_{JI;M}\rangle$ are shown in Fig. \ref{chirband}. As mentioned before, they are denoted as $T1, T2, T3$ and $T4$, respectively. The strength of the $qQ$ interaction, $X^{h}_{2}$, was chosen so that the energy levels of the band $ T1$ be as close as possible to those of the experimental dipole band D4. The quasiparticle energy for 
$j=h_{11/2}$ is taken equal to 2 MeV, which for a single $j$ calculation would correspond to a paring strength G=0.67 MeV. The agreement with the data is improved by switching on the anharmonic particle-core interaction. Since the monopole-monopole term determines just a constant shift for the energy levels, we ignore it. As for the quadrupole-quadrupole and hexadecapole-hexadecapole interactions, we considered equal strengths but of different sign i.e., $X^{an}_{2}=-X^{an}_{4}$. This choice quenches the rotation influence on energies in the upper part of the band. Switching on the $sS$ interaction,
the degeneracy of the four bands, $T1, T2, T3$ and $T4$, is lifted up. The strength of the $sS$ interaction is determined such that the distance between two chiral partner bands equals  the corresponding experimental data.  The parameters involved in the model calculations and determined as mentioned above are collected in Table1. 
\begin{table}
\begin{tabular}{|cccccccccccc|}
\hline
   $\rho$  &   $A_1$   &     $A_2$    &   $A_3$    &   $A_4$ &  $X^{h}_{2}$  &  $X_{sS}$ &$X^{an}_2$&$X^{an}_4$ & $g_p$[$\mu_N$]  &  $g_n$[$\mu_N$] &  $g_F$[$\mu_N$] 
\\
\hline
1.6   &   1.114   &-0.566    &   4.670  &  0.0165  &  0.200  &  0.02 &-0.4&0.4& 0.492 & 0.377 &1.289\\  
\hline
\end{tabular}
\caption{The structure coefficients involved in the model Hamiltonian and described as explained in the text, are given in units of MeV. We also list the values of the deformation parameter
$\rho$ (a-dimensional) and the gyromagnetic factors of the three components, protons, neutrons and fermions, given in units of nuclear magneton ($\mu_N$).}
\end{table}
\newpage
\begin{figure}[h!]
\includegraphics[width=0.5\textwidth]{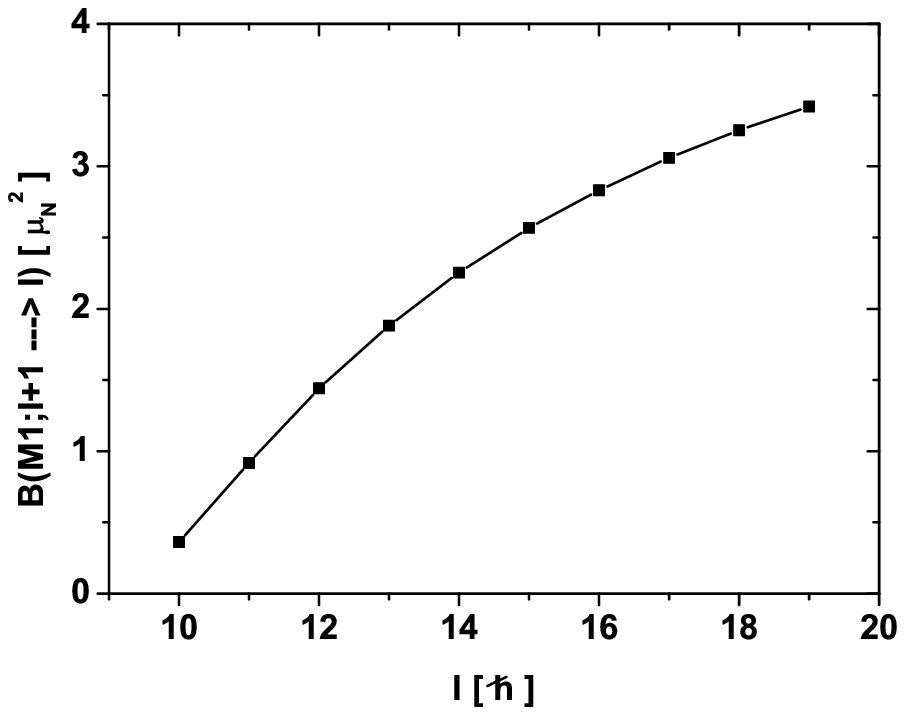}\includegraphics[width=0.5\textwidth]{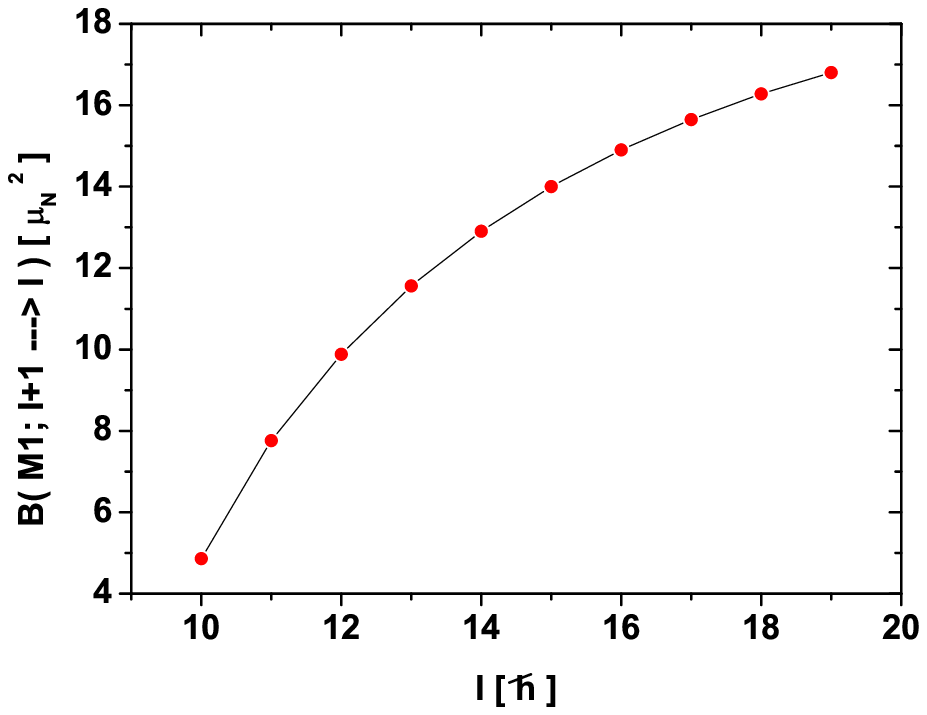}
\includegraphics[width=0.5\textwidth]{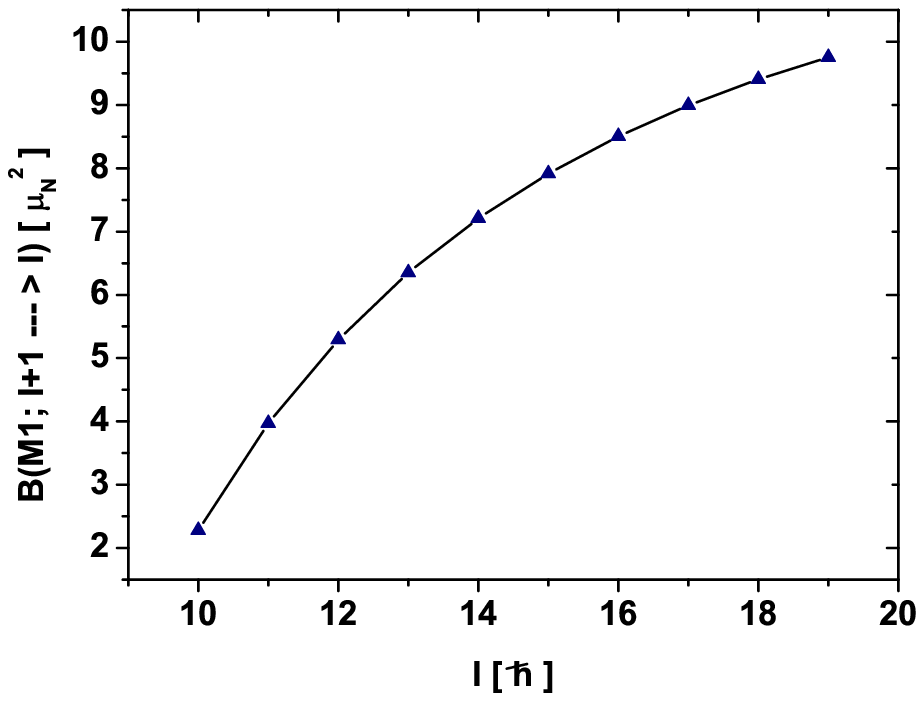}\includegraphics[width=0.5\textwidth]{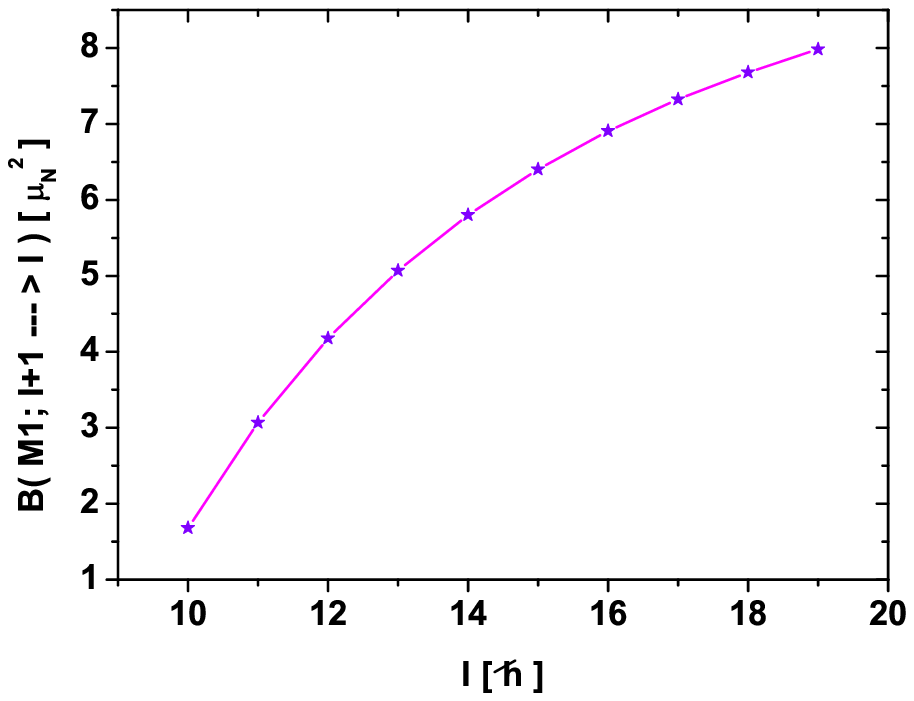}        
\caption{\scriptsize{Upper left panel: The B(M1) values for the transitions  $I+1\to I$ within the chiral $T_1$ band.
Upper right panel: The B(M1) values for the transitions  $I+1\to I$ within the chiral $T_2$ band.
Bottom left panel: The B(M1) values for the transitions  $I+1\to I$ within the chiral $T_3$ band. Bottom right panel: The B(M1) values for the transitions $I+1\to I$ within the chiral $T_4$ band.}}
\label{chirM1}
\end{figure}
We note that the calculated energies for the band $T1$ agree reasonable well with those of the experimental band $D4$, except for the state $10^+$ which is closer to  
$10^+$ of the ground band.

The experimental dipole band $D4$ was not well understood in Ref. [10] in the framework of Cranked Nilson Strutinsky $(CNS)$ and Tilted Axis Cranking $(TAC)$ calculations. The two proposed configurations involve either two positive-parity proton orbitals from the $(d_{5/2}, g_{7/2})$ sub-shell, the second and fourth above the Fermi level, or four orbitals - two proton and  two neutron orbitals of opposite parity. These configurations are calculated by the $TAC$ model at excitation energies relative to the yrast band $L1$ much higher than the experimental one (more than 0.5 MeV), and are therefore questionable, since band $D4$ is the lowest excited dipole band with band-head spin around $10^+$, for which one would expect a better agreement between experiment and theory. On the other hand, in Ref. [10] only the prolate deformed configurations were investigated, in which the particle-like proton $h_{11/2}^2$ configuration is favored. In the present calculations a hole-like proton $h_{11/2}^2$ configuration is assumed, with the quasiparticle energy for $j=h_{11/2}$ equal to 2 MeV, which for a single $j$ calculation would correspond to  pairing strength $G=0.67$ MeV.

Remarkable the fact that there are two experimental levels which might be associated to two states of the calculated chiral partner band $T2$. 
The state $11^+$ at 4.381 MeV has been reported in Ref. \cite{Petra1}, while the tentative $(12^+)$ state at 4.737 MeV has been identified after revisiting the same experimental data reported in 
Ref. \cite{Petra1}. The new $(12^+)$ state is populated by a weak transition of 332 keV from the $13^+$ state of band D4 and decays to the $11^+$ state at 4.381 MeV through a 356-keV transition. As the new $(12^+)$ state is very weakly populated, one could not assign a definite spin-parity. However, the $12^+$ assignment is the most plausible, since other spin values or negative parity would led to unrealistic values of the connecting transitions. The corresponding $11^+$ and $12^+$ calculated levels of band $T2$ have energies of 4.776 and 4.324 MeV, respectively, which are very close to the experimental values of 4.737 and 4.381 MeV.

The magnetic dipole transitions are calculated with the operator:
\begin{equation}
{\cal M}_{1,m}=\sqrt{\frac{3}{4\pi}}\left(g_pJ_{p,m}+g_nJ_{n,m}+g_FJ_{F,m}\right).
\end{equation}
The collective proton and neutron gyromagnetic factors were calculated as explained in Ref. \cite{AAR2014}. For the sake of completeness we give the results also here:
\begin{eqnarray}
\left(\begin{matrix}g_p \cr g_n \end{matrix}\right) = \frac{3ZR_0^2}{8\pi k_p^2}\frac{Mc^2}{(\hbar c)^2}\left(\begin{matrix}A_1+6A_4\cr \frac{1}{5}A_3\end{matrix}\right)
\end{eqnarray}
The same notations as in Ref. \cite{AAR2014} were used. The results for the collective gyromagnetic factors are those shown in Table 1.

The magnetic dipole reduced transition probabilities within each of the four chiral bands are plotted in Fig. \ref{chirM1} as function of the final state angular momentum.
Note that although the B(M1) values, in the four chiral bands, have similar behavior as function of the angular momentum, quantitatively they substantially differ from each other. Remarkable is the large magnitude of these transitions within the band $T2$ where the angular momentum of the fermions is oriented differently than that in band $T1$. It seems that changing the sign of one gyromagnetic factor favor the increase of the B(M1) values. We also notice that although the bands $T3$ and $T4$ are degenerate, the associated B(M1) values are different.
Concluding, the dependence of the magnetic dipole transition intensities on the nature of the chiral band is a specific feature of the present formalism. 

The effect of the chiral transformation on the B(M1) values can be understood by analyzing the relative signs of the collective and fermionic transition amplitudes.
Indeed, with the results presented in Appendix A, the reduced transition probability can be written as:
\begin{equation}
B(M1;I+1\to I)=\frac{3}{4\pi}\left(A^{(I)}_{pn}+ A^{(I)}_F\right)^2,
\end{equation}
where $A^{(I)}_{pn}$  denotes the terms of the transition matrix elements which are linear combination of the gyromagnetic factors $g_p$ and $g_n$,
while $A^{(I)}_F$ is that part which is proportional to $g_F$. Their values for the transitions in the four chiral bands are listed in Table II.
\begin{table}
\begin{tabular}{|c|cc|cc|cc|cc|}
\hline
   &\multicolumn{2}{c|}{$T_1$ band}&\multicolumn{2}{c|}{$T_2$ band}&\multicolumn{2}{c|}{$T_3$ band}&\multicolumn{2}{c|}{$T_4$ band}\\
\hline
I  &$A^{(I)}_{pn}$&$A^{(I)}_F$&$A^{(I)}_{pn}$&$A^{(I)}_F$&$A^{(I)}_{pn}$&$A^{(I)}_F$&$A^{(I)}_{pn}$&$A^{(I)}_F$\\
\hline
$10^+$&-1.639&2.871&-1.639&-2.871&0.217&2.871&-0.217&2.871\\
$11^+$&-1.870&3.831&-1.870&-3.831&0.247&3.831&-0.247&3.831\\
$12^+$&-1.989&4.445&-1.989&-4.445&0.263&4.445&-0.263&4.445\\
$13^+$&-2.076&4.883&-2.076&-4.883&0.275&4.883&-0.275&4.883\\
$14^+$&-2.140&5.213&-2.140&-5.213&0.283&5.213&-0.283&5.213\\
$15^+$&-2.189&5.469&-2.189&-5.469&0.290&5.469&-0.290&5.469\\
$16^+$&-2.228&5.673&-2.228&-5.673&0.295&5.673&-0.295&5.673\\
$17^+$&-2.259&5.838&-2.259&-5.838&0.299&5.838&-0.299&5.838\\
$18^+$&-2.283&5.974&-2.283&-5.974&0.302&5.974&-0.302&5.974\\
$19^+$&-2.303&6.087&-2.303&-6.087&0.305&6.087&-0.305&6.087\\
\hline
\end{tabular}
\caption{The magnetic dipole proton-neutron and fermion transition amplitudes for the bands $T1, T2, T3, T4.$}
\end{table}

We note that for the bands $T2$ and $T3$ the contributions of the collective and fermion transition amplitudes are in phase,  while for the other bands they exhibit different phases, which
 explains the large magnitude of the  transitions in the bands $T2$ and $T3$ compared  with those within the bands $T1$ and $T4$. In fact, this is a proof that by a chiral transformation
the transversal magnetic moment is increased. 
\begin{figure}[h!]
\begin{center}
\includegraphics[width=0.5\textwidth]{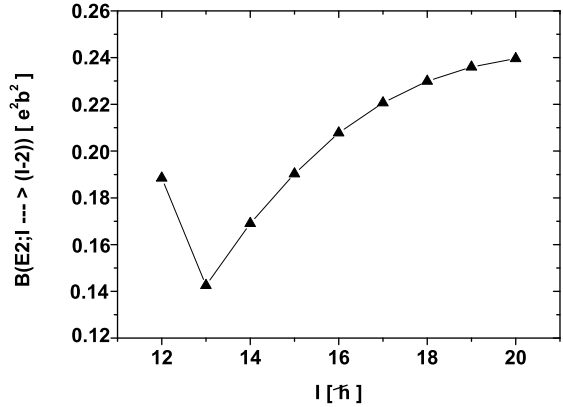}   
\end{center}
\caption{\scriptsize{The B(E2) values for the intra-band transitions.}}
\label{FigBE2}
\end{figure}
\begin{figure}[h!]
\begin{center}
\includegraphics[width=1.1\textwidth]{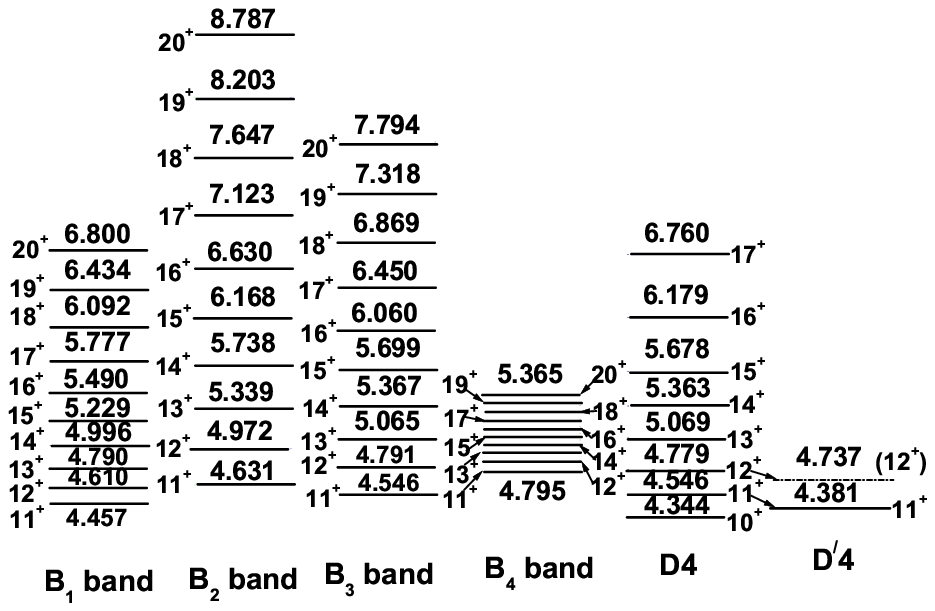}     
\end{center}
\caption{\scriptsize{ The excitation energies, given in MeV, for the bands $B_1, B_2, B_3$ and  $B_4$. The experimental chiral partner bands $D4$ and $D^{\prime}4$ are also shown. The band $B_3$ is to be compared with the experimental band D4.}}
\label{chirdip}
\end{figure}

The quadrupole electric transition probabilities were calculated using for the transition operator the expression:
\begin{equation}
{\cal M}_{2\mu}=\frac{3ZeR^2_0}{4\pi}\alpha_{p\mu}
\end{equation}
where the quadrupole collective coordinate, $\alpha_{p\mu}$, is related with the corresponding boson operators by the canonical transformation:
\begin{equation}
\alpha_{p\mu}=\frac{1}{\sqrt{2}k_p}(b_{p\mu}+(-)^{\mu}b_{p,-\mu}).
\end{equation}
The standard notations for the nuclear charge, electron charge and nuclear radius are used.
The above transformation is canonical irrespective the value of $k_p$, which is determined as described in Ref. \cite{AAR2015}, with the result:
\begin{equation}
k_{p}^{2}=\frac{3}{16\pi}AR_0^2\frac{Mc^2}{(\hbar c)^2}\left(A_1+6A_4+\frac{1}{5}A_3\right).
\end{equation}
Here $M$ denotes the nucleons mass, $c$ the light velocity, while $A_1, A_3, A_4$ are the structure coefficients involved in the phenomenological GCSM Hamiltonian.
Since the quadrupole transition operator is invariant to any chiral transformation, the B(E2) values  for the intra-band transitions in the four chiral bands are the same.
The common values are represented as function of the angular momentum in Fig. \ref{FigBE2}.
 We notice the small B(E2) transition probabilities which, in fact, is a specific feature of the chiral bands.

\subsection{Chiral bands with $H_2$($H_3$), $|\Psi^{(2qp;J1)}_{JI;M}\rangle$ and chiral transformed states}

Here we attempt to describe the experimental dipole band $D_4$ by using the states from the basis (\ref{basis2}) and the Hamiltonians:

\begin{eqnarray}
H_2&=&H^{\prime}_{GCSM}+2A_4{\bf J_p}\cdot {\bf J_n}+\sum_{\alpha}E_{a}a^{\dagger}_{\alpha}a_{\alpha} -X_{sS}{\bf {J}_F}\cdot{\bf {J}_c},\\
\label{HHprimSS}
H_3&=&H^{\prime}_{GCSM}+2A_4{\bf J_p}\cdot {\bf J_n}+\sum_{\alpha}E_{a}a^{\dagger}_{\alpha}a_{\alpha}.
\label{H2Hprim}
\end{eqnarray}
The results are presented in Fig. 12.  
The bands $B_1$ and $B_2$ were obtained by averaging $H_2$ (\ref{HHprimSS}) with the functions $|\Psi^{(2qp;J1)}_{JI;M}\rangle$, $C_{12}|\Psi^{(2qp;J1)}_{JI;M}\rangle$ respectively, while the bands
$B_3$ and $B_4$ by averaging $H_3$ with the functions  $|\Psi^{(2qp;J1)}_{JI;M}\rangle$, $C_{13}|\Psi^{(2qp;J1)}_{JI;M}\rangle$, respectively. The two Hamiltonians  differ by the spin-spin interaction which is missing in $H_3$. The reason for which one uses two different Hamiltonians consists of the fact that for the band $B_4$ the spin-spin interaction is ineffective. On the other hand it is desirable that the two partner bands $B_3$ and $B_4$ be described by a sole Hamiltonian. Again, the band obtained by averaging $H_3$ with the function 
$C_{14}|\Psi^{(2qp;J1)}_{JI;M}\rangle$ is identical with $B_4$. All parameters involved in the two Hamiltonians are the same as those from Table 1, except for the quasiparticle energy which here is taken equal to 1.461 MeV. This corresponds, for a single $j$ shell, to a paring strength G=0.238 MeV.         
It is remarkable the excellent agreement between the excitation energies of the band $B_3$ and the corresponding experimental ones given in the first column of Fig. \ref{chirband}. Practically, the
experimental excitation energies are obtained by shifting the core weighted energies with the two quasiparticle energy. The energy spacing in the partner band $B_4$ is constant (about 60 keV) with a deviation of at most 3 keV.

\begin{table}
\begin{tabular}{|c|ccc|ccc|ccc|ccc|}
\hline
   &\multicolumn{3}{c|}{$B_1$ band}&\multicolumn{3}{c|}{$B_2$ band}&\multicolumn{3}{c|}{$B_3$ band}&\multicolumn{3}{c|}{$B_4$ band}\\
\hline
I  &B(M1)&$A^{(I)}_{pn}$&$A^{(I)}_F$&B(M1)&$A^{(I)}_{pn}$&$A^{(I)}_F$&B(M1)&$A^{(I)}_{pn}$&$A^{(I)}_F$&B(M1)&$A^{(I)}_{pn}$&$A^{(I)}_F$\\
\hline
$11^+$&0.662&-1.041&2.705 &3.352 &-1.041& -2.705&1.931& 0.138& 2.705& 1.574&-0.138& 2.705   \\
$12^+$&1.664&-0.989&3.629 &5.093 & -0.989&-3.629&3.376& 0.131& 3.629& 2.922& -0.131& 3.629    \\
$13^+$&2.596&-0.933&4.231 &6.365 &-0.933& -4.231&4.526& 0.123& 4.231& 4.027& -0.123& 4.231    \\
$14^+$&3.409&-0.886&4.665 &7.358 & -0.886&-4.665&5.460& 0.117& 4.665& 4.938& -0.117& 4.665    \\
$15^+$&4.109&-0.847&4.996 &8.151 & -0.847&-4.996&6.229& 0.112& 4.996& 5.694& -0.112& 4.996   \\
$16^+$&4.711&-0.813&5.256 &8.796 &-0.813& -5.256&6.868& 0.108& 5.256& 6.328& -0.108& 5.256     \\
$17^+$&5.231&-0.784&5.465 &9.325 &-0.784& -5.465&7.405& 0.104& 5.465& 6.863& -0.104& 5.465     \\
$18^+$&5.681&-0.758&5.637 &9.762 & -0.758&-5.637&7.857& 0.100& 5.637& 7.317& -0.100& 5.637     \\
$19^+$&6.071&-0.734&5.777 &10.123& -0.734&-5.777&8.239& 0.097& 5.777& 7.702& -0.097& 5.777      \\
\hline
\end{tabular}
\caption{The magnetic dipole proton-neutron and fermion transition amplitudes for the bands $B_1, B_2, B_3, B_4.$}
\end{table}

It is very interesting to note that under the chiral transformation $C_{13}$ the rotational term $\hat{J}^2_c$ involved in $H_{GCSM}$ becomes $(J_p-J_n)^2$. This term appearing in the chiral transformed $H_3$ is essential in determining the partner band $B_4$. On the other hand we recall that such a term is used by the two rotor model to define the scissors mode. In that respect the partner band $B_4$ may be interpreted as the second order scissors band.

\section{Conclusions}
In the previous sections we described a formalism for the chiral bands. The application is made for the isotope of $^{138}$Nd where some experimental data are available \cite{Petra1,HJLi}.
The phenomenological core is described by the GCSM. The single particles move in a shell model mean-field and alike nucleons interact among themselves through
pairing. The model Hamiltonian was treated within a restricted space associated to the phenomenological core and the   subspace of aligned two proton quasiparticle states coupled to the  states of the phenomenological core states. Two scenarios for the particle-core Hamiltonian as well as for the particle-core states are considered: 

{\bf a)} The particle-core Hamiltonian involves as  phenomenological factor a linear quadrupole as well as an anharmonic boson quadrupole and an anharmonic boson hexadecapole term. Also a spin-spin interaction is taken into consideration.  The model Hamiltonian was treated within a restricted space associated to the phenomenological core and the   subspace of two aligned proton quasiparticle states coupled with the  states of the phenomenological ground band. The former space is spanned by the model projected states assigned to six bands: ground (g), $\beta,\gamma,\widetilde{\gamma},1^+,\widetilde{1^+}$. This basis is enlarged by adding the chiral transformed states. The eigenvalues of the model Hamiltonian within this restricted basis are arranged in four bands denoted by $T1, T2, T3$ and $T4$, respectively.
The experimental band D4 is suspected to be of chiral nature. This  is compared with the theoretical band $T1$. The bands $T2, T3,T4$  are also studied. An experimental chiral partner for band D4
is also proposed, which is associated with the calculated band $T2$.
We calculated not only the energies of these bands but also the intra-band M1 and E2 transitions. Since the M1 transition operator does not commute with the chiral transformations relating the right and left-handed intrinsic reference frames, the B(M1) values are changed when we pass from one band to another. On the contrary, the E2 transition operator is invariant to the mentioned chiral transformations and therefore B(E2) values are the same in all four chiral bands. It is remarkable that although the bands $T3$ and $T4$ are degenerate, they are characterized by different intra-band $M1$ transitions.
Both the M1 and E2 transitions exhibit the properties which are specific to the chiral bands. Indeed, the B(M1) values are large, while the  B(E2) values are small.

{\bf b)} The  phenomenological core is described by the GCSM Hamiltonian, as in the case denoted by a). To this one adds the independent quasiparticle term and the spin-spin interaction. The basis used for treating this Hamiltonian  consists of the core states, the six bands, and   two aligned proton quasiparticles, from the $j=h_{11/2}$ sub-shell, coupled with the phenomenological magnetic dipole  states describing the core. Also the chiral transformed states are members of the model  basis. The resulting bands are denoted by $B_1, B_2, B_3$ and $B_4$, respectively. Energies and 
$M1$ properties for  these bands are quantitatively studied. Note that for describing the bands $B_3$ and $B_4$ there is only one parameter, that is the two quasiparticle energy of the single $j$
state $h_{11/2}$, which shift all energies in the two bands. The energies of band $B_3$ agree very well with the corresponding experimental data. The band head  of the B-like bands is the state $11^+$. In this scenario the head state for the experimental band D4, $10^+$, belongs to the ground band of the core. Note that the effect of the chiral transformation $C_{12}$ on the energies of band $B_1$ is to increase them, while the transformation $C_{13}$ applied on $B_3$ compresses the energy levels.

In both scenarios the energy spacings in all four bands is almost constant. Also the intra-band B(M1) values are large and different when one passes from one band to another.
On the contrary, the B(E2) values do not depend on band and in general are small. As a matter of fact these properties confer the bands a chiral character.

Experimental data for energies of the chiral partner bands as well as for electromagnetic intra-band transitions would be a stringent test for the hypothesis advanced in the present paper.
Our work proves that the mechanism for chiral symmetry breaking, which also favors a large transverse component for the dipole magnetic transition operator \cite{Frau}, is not unique.

The two bases were chosen since the core states are low in energy and therefore  are the most favored ones. However, the results described above show that the basis with the dipole collective states
involved provides a more realistic description of the experimental band D4. Moreover the partner band $B_4$ may be interpreted as the second order scissors band. 

It is clear that an improvement of the two descriptions may be obtained by diagonalizing the model Hamiltonian consisting of $H_{GCSM}$ plus the particle-core coupling terms in a larger basis obtained by the reunion of the two bases (\ref{basis1}) and (\ref{basis2})


For odd-odd nuclei several groups identified twin bands in medium mass regions \cite{Petrache96,Simon2005,Vaman,Petrache06}  and even for heavy mass regions \cite{Balab,Frau97,Dimi}.
The formalisms proposed are based either on the Tilted Axis Cranking (TAC) approach \cite{Frau00} or on the two quasiparticle-triaxial rotor coupling model \cite{Fas1,Fas2,Fas3,Fas4}.
Although the efforts were mainly focused on identifying and describing the chiral twin bands in odd-odd nuclei, few results for even-odd \cite{22,23,24,25,26,Aya} and even-even \cite{27} nuclei were also reported.

Finally, we mention again that the formalism proposed in the present paper concerns the even-even nuclei and is based on a new concept. Indeed, there are few features which contrasts the main characteristics of the elegant model proposed by Frauendorf for odd-odd nuclei \cite{Frau00,Frau}.
Indeed, here the right- and left-handed frames are the angular momentum carried by two aligned protons and by proton and neutron bosons respectively, associated to the core.
Within the model proposed by Frauendorf the shears motion is achieved by  one proton-particle and one neutron-hole, while here the shears blades are the proton and neutron components of the core.
The B(M1) values are maximal at the beginning of the band and decrease with angular momentum and finally, when the shears are closed, they are vanishing since there is no transverse magnetic momentum any longer. By contrast, here the B(M1) value is an increasing function of the angular momentum. In both models the dominant contribution to the dipole magnetic transition probability
is coming from particles sub-system. This property is determined by the specific magnitudes of the  gyromagnetic factors associated to the three components of the system.
Since the two schematic models reveal  some complementary magnetic properties of nuclei they might cover different areas of nuclear spectra.

{\bf Acknowledgment.} This work was supported by the Romanian Ministry for Education Research Youth and Sport through the CNCSIS project ID-2/5.10.2011.

\section{Appendix A}
Here we give the analytical expressions for the matrix elements of the magnetic dipole transition operator corresponding to the chiral band states.

\begin{eqnarray}
&&\langle \Psi^{(2qp;c)}_{JI}||M_1||\Psi^{(2qp;c)}_{JI^{\prime}}\rangle = 2\sqrt{\frac{3}{4\pi}}N^{(2qp;c)}_{JI}N^{(2qp;c)}_{JI^{\prime}}\hat{I^{\prime}}
\sum_{J_1}\left(N^{(g)}_{J_{1}}\right)^{-2}C^{J J_1 I}_{J 0 J}C^{J J_1 I^{\prime}}_{J 0 J}\\
&\times&\left[\hat{J_1}W(J J_1 I 1;I^{\prime}J^{\prime})(g_p+g_n)\sqrt{\widetilde{J}_{pJ_1}(\widetilde{J}_{pJ_1}+1)}+
             \hat{J}W(I^{\prime}J^{\prime}1J;JI)g_F\sqrt{J(J+1)}\right].\nonumber
\end{eqnarray}

The matrix element between the transformed states are:
\begin{eqnarray}
&&\hspace*{-1cm}\langle \Psi^{(2qp;c)}_{JI}C_{12}^{\dagger}||M_1||C_{12}\Psi^{(2qp;c)}_{JI^{\prime}}\rangle = 2\sqrt{\frac{3}{4\pi}}N^{(2qp;c)}_{JI}N^{(2qp;c)}_{JI^{\prime}}\hat{I^{\prime}}
\sum_{J_1}\left(N^{(g)}_{J_{1}}\right)^{-2}C^{J J_1 I}_{J 0 J}C^{J J_1 I^{\prime}}_{J 0 J}\\
&&\hspace*{-1cm}\times\left[\hat{J_1}W(J J_1 I 1;I^{\prime}J^{\prime})(g_p+g_n)\sqrt{\widetilde{J}_{pJ_1}(\widetilde{J}_{pJ_1}+1)}-
             \hat{J}W(I^{\prime}J^{\prime}1J;JI)g_F\sqrt{J(J+1)}\right].\nonumber
\end{eqnarray}

\begin{eqnarray}
&&\hspace*{-1cm}\langle \Psi^{(2qp;c)}_{JI}C_{13}^{\dagger}||M_1||C_{13}\Psi^{(2qp;c)}_{JI^{\prime}}\rangle = 2\sqrt{\frac{3}{4\pi}}N^{(2qp;c)}_{JI}N^{(2qp;c)}_{JI^{\prime}}\hat{I^{\prime}}
\sum_{J_1}\left(N^{(g)}_{J_{1}}\right)^{-2}C^{J J_1 I}_{J 0 J}C^{J J_1 I^{\prime}}_{J 0 J}\\
&&\hspace*{-1cm}\times\left[\hat{J_1}W(J J_1 I 1;I^{\prime}J^{\prime})(-g_p+g_n)\sqrt{\widetilde{J}_{pJ_1}(\widetilde{J}_{pJ_1}+1)}+
             \hat{J}W(I^{\prime}J^{\prime}1J;JI)g_F\sqrt{J(J+1)}\right].\nonumber
\end{eqnarray}
\begin{eqnarray}
&&\hspace*{-1cm}\langle \Psi^{(2qp;c)}_{JI}C_{14}^{\dagger}||M_1||C_{14}\Psi^{(2qp;c)}_{JI^{\prime}}\rangle = 2\sqrt{\frac{3}{4\pi}}N^{(2qp;c)}_{JI}N^{(2qp;c)}_{JI^{\prime}}\hat{I^{\prime}}
\sum_{J_1}\left(N^{(g)}_{J_{1}}\right)^{-2}C^{J J_1 I}_{J 0 J}C^{J J_1 I^{\prime}}_{J 0 J}\\
&&\hspace*{-1cm}\times\left[\hat{J_1}W(J J_1 I 1;I^{\prime}J^{\prime})(g_p-g_n)\sqrt{\widetilde{J}_{pJ_1}(\widetilde{J}_{pJ_1}+1)}+
             \hat{J}W(I^{\prime}J^{\prime}1J;JI)g_F\sqrt{J(J+1)}\right].\nonumber
\end{eqnarray}
The reduced matrix elements for the M1 transition operator in the basis (\ref{basis2}) have been analytically derived in Ref.\cite{AAR2014}.
The electric quadrupole transition probabilities are determined by the matrix elements:
\begin{eqnarray}
\langle \Psi^{(2qp;c)}_{JI}||b^{\dagger}_2+b_2||\Psi^{(2qp;c)}_{JI^{\prime}}\rangle & =& N^{(2qp;c)}_{JI}N^{(2qp;c)}_{JI^{\prime}}2d\hat{I^{\prime}}
\sum_{J_1J_2}\hat{J}_1C^{J J_1 I}_{J 0 J}C^{J J_2 I^{\prime}}_{J 0 J}C^{J_2 2 J_1}_{0\;\; 0\;\; 0}W(J J_2 I 2; I^{\prime}J_1)\nonumber\\
&\times&\left[\frac{2J_2+1}{2J_1+1}\left(N^{(g)}_{J_1}\right)^{-2}+\left(N^{(g)}_{J_2}\right)^{-2}\right].
\end{eqnarray}
Similar expressions are obtainable also for the basis (\ref{basis2}).


\begin{references}
\bibitem{LoIu1} N. Lo Iudice and F. Palumbo, Phys. Rev. Lett. {\bf 41}, 1532 (1978).
\bibitem{LoIu2} G. De Francheschi, F. Palumbo and N. Lo Iudice, Phys. Rev. {\bf C29}, 1496 (1984).
\bibitem{LoIu3} N. Lo Iudice, Phys. Part. Nucl. {\bf 25 }, 556, (1997).
\bibitem{Zawischa} D. Zawischa, J. Phys. G{\bf 24}, 683, (1998).
\bibitem{Frau}S. Frauendorf, Rev. Mod. Phys.  {\bf 73}, 463 (2001).
\bibitem{Jenkins} D. G. Jenkins et al., Phys. Rev. Lett. {\bf 83}, 500 (1999).
\bibitem{AAR2014} A. A. Raduta, C. M. Raduta and A. Faessler, J. Phys. G: Nucl. Part. Phys {\bf 41}, 035105 (2014).
\bibitem{AAR2015} A. A. Raduta and C. M. Raduta J. Phys. G: Nucl. Part. Phys {\bf 42}, 065105 (2015).
\bibitem{RB2014} A. A. Raduta and R. Budaca, Ann. Phys. (NY) {\bf 347} (20140 141.
\bibitem{Lia90} Y. Liang, R. Ma, E. S. Paul, N. Xu, D. B. Fossan, J. Y. Zhang and F. Donau, Phys. Rev. Lett. {\bf 64}, (29 (1990).
\bibitem{Paul90}E. S. Paul, D. B. Fossan, Y. Liang, R. Ma, N. Xu, R. Wadsworth, I. Jenkins, P. J. Nolan, Phys. Rev. {\bf C 41}1576 (1990).
\bibitem{Rad2}A. A. Raduta, A. Faessler and V. Ceausescu, Phys. Rev. {\bf C 36}, 439 (1987).
\bibitem{Rad1}A. A. Raduta {\it et al.}, Phys. Lett. {\bf B 121}1; Nucl. Phys. {\bf A 381}, 253 (1982).
\bibitem{Sheline} R. K. Sheline, Rev. Mod. Phys. {\bf 32}, 1, (1960); M. Sakai, Nucl. Phys. {\bf A 104}, 301 (1976). 
\bibitem{Rad3}A. A. Raduta, I. I. Ursu and D. S. Delion, Nucl. Phys. {\bf A 475}, 439 (1987).
\bibitem{Rad4}A. A. Raduta and D. S. Delion, Nucl. Phys. {\bf A 491}, 24 (1989).
\bibitem{4} N. Lo Iudice, {\it et al.}, Phys. Lett. {\bf B 300} (1993) 195; Phys. Rev. {\bf C 50}, 127 (1994).
\bibitem{Lima}A. A. Raduta, C. Lima and Amand Faessler, Z. Phys. A - Atoms and Nuclei {\bf 313},  69 (1983).
\bibitem{4} N. Lo Iudice, {\it et al.}, Phys. Lett. {\bf B 300} (1993) 195; Phys. Rev. {\bf C 50}, 127 (1994).
\bibitem{Iud}N. Lo Iudice, A. A. Raduta and D. S. Delion, Phys. Rev. {\bf C50}, 127 (1994).
\bibitem{Grei} V. Maruhn-Rezwani, W. Greiner and J. A. Maruhn, Phys. Lett. {\bf 57 B}, 109 (1975).
\bibitem{Novos} A. Novoselski and I. Talmi, Phys. Lett. {\bf 60 B}, 13 (1985).
\bibitem{Petra1} C. M. Petrache {\it et al}, Phys. Rev. {\bf C 86}, 044321 (2012).
\bibitem{HJLi} H. J. Li {\it et al}, Phys. Rev. {\bf C 87}, 057303 (2013).
\bibitem{AARBook} A. A. Raduta, {\it Nuclear Structure with Coherent States}, Springer, ISBN 978-3-319-14641-6, Cham Heidelberg New York Dordrecht London. 
\bibitem{Grod} E. Grodner, Acta Physica Polonica {\bf B 39}, No. 2, 531 (2008).
\bibitem{Desla} J. Deslauriers, S. C. Gujrathi, S. K. Mark, Z. Phys. {\bf A 303}, 151 (1981).
\bibitem{Petrache96}C. M. Petrache {\it et al.,} Nucl. Phys. {\bf A597}, 106 (1996).
\bibitem{Simon2005}A. J. Simon {\it et al.,} Jour. Phys. G: Nucl. Part. Phys {\bf 31}, 541 (2005).
\bibitem{Vaman}C. Vaman, {\it et al.}, Phys. Rev. Lett. {\bf 92}, 032501 (2004).
\bibitem{Petrache06}C. M. Petrache, {\it et al.}, Phys. Rev. Lett. {\bf 96}, 112502 (2006).
\bibitem{Balab} D. L. Balabanski, {\it et al.,} Phys. Rev. C {\bf 70}, 044305 (2004).
\bibitem{Frau97}S. Frauendorf and J. Meng, Nucl. Phys. {\bf A 617}, 131 (1997).
\bibitem{Dimi} V. I. Dimitrov, S. Frauendorf and F. D\:{o}nau, Phys. Rev. Let. {\bf 84}, 5732 (2000).
\bibitem{Frau00} S. Frauendorf, Nuclear Physics {\bf A677}, 115 (2000)
\bibitem{Fas1} H. Toki and Amand Faessler, Nucl. Phys. {\bf A 253}, 231 (1975).
\bibitem{Fas2} H. Toki and Amand Faessler, Z. Phys.. {\bf A 276},  35 (1976).
\bibitem{Fas3} H. Toki and Amand Faessler, Phys. Lett. {\bf B 63} (1976) 121.
\bibitem{Fas4} H. Toki, H. L. Yadav and Amand Faessler, Phys. Lett. {\bf B 66},  310 (1977).
\bibitem{22} S. Zhu {\it et al.} Phys. Rev. Lett. {\bf 91 }, 132501 (2003)
\bibitem{23} J. A. Alcantara-Nunez {\it et al.}, Phys. Rev. {\bf C69} 024317 (2004).
\bibitem{24} J. Timar {\it et al.}, Phys. Lett. {\bf B 598}, 178 (2004). 
\bibitem{25} J. Timar {\it et al.}, Phys. Rev. {\bf C 73}, 011301 (R) (2006).
\bibitem{26} Y. X. Luo, {\it et al.}, Chin. Phys. Lett. {\bf B 26}, 082301 (2009).
\bibitem{Aya} A. D. Ayangeakaa {\it et al.}, Phys. Rev. Lett. {\bf 110}, 172504 (2013). 
\bibitem{27} E. Mergel {\it et al.}, Eur. Phys. J. {\bf A 15}, 417 (2002).
\end{references}
\end{document}